\documentclass[iop,numberedappendix]{emulateapj}

\usepackage[backref,breaklinks,colorlinks,citecolor=blue]{hyperref} 
\usepackage{amsmath}

\newcommand{\gps}{\ensuremath{g_{\rm P1}}}
\newcommand{\rps}{\ensuremath{r_{\rm P1}}}
\newcommand{\ips}{\ensuremath{i_{\rm P1}}}
\newcommand{\ipsl}{\ensuremath{i_{\rm P1,lim}}}

\newcommand{\zps}{\ensuremath{z_{\rm P1}}}

\newcommand{\zext}{\ensuremath{z_{\rm P1,ext}}}

\newcommand{\yps}{\ensuremath{y_{\rm P1}}}

\newcommand{\yext}{\ensuremath{y_{\rm P1,ext}}}

\newcommand{\PS}{\protect \hbox {Pan-STARRS1}}

\newcommand{\ggrond}{\ensuremath{g_{\rm G}}}
\newcommand{\rgrond}{\ensuremath{r_{\rm G}}}
\newcommand{\igrond}{\ensuremath{i_{\rm G}}}
\newcommand{\zgrond}{\ensuremath{z_{\rm G}}}
\newcommand{\Jgrond}{\ensuremath{J_{\rm G}}}
\newcommand{\Hgrond}{\ensuremath{H_{\rm G}}}
\newcommand{\Kgrond}{\ensuremath{K_{\rm G}}}

\newcommand{\intt}{\ensuremath{I_{\rm E}}}
\newcommand{\zntt}{\ensuremath{Z_{\rm E}}}
\newcommand{\Jntt}{\ensuremath{J_{\rm S}}}

\newcommand{\zotk}{\ensuremath{z_{\rm O2K}}}
\newcommand{\Yotk}{\ensuremath{Y_{\rm O2K}}}
\newcommand{\Jotk}{\ensuremath{J_{\rm O2K}}}
\newcommand{\Hotk}{\ensuremath{H_{\rm O2K}}}

\newcommand{\ydp}{\ensuremath{Y_{\rm R}}}
\newcommand{\jdp}{\ensuremath{J_{\rm R}}}

\newcommand{\Icafos}{\ensuremath{i_{\rm C}}}

\newcommand{\Jtmass}{\ensuremath{J_{\rm 2M}}}
\newcommand{\Htmass}{\ensuremath{H_{\rm 2M}}}
\newcommand{\Ktmass}{\ensuremath{K_{\rm 2M}}}

\newcommand{\lya}{\mathrm{Ly}\ensuremath{\alpha}}
\newcommand{\nv}{\mathrm{N}\ensuremath{\,\textsc{v}}}
\newcommand{\lyaew}{\mathrm{EW}(\lya + \nv)}

\newcommand{\mgii}{Mg\ensuremath{\,\textsc{ii}}}

\newcommand{\civ}{C\ensuremath{\,\textsc{iv}}}

\newcommand{\cii}{[\ion{C}{2}]}

\newcommand{\siivpoiv}{Si\ensuremath{\,\textsc{iv}\,+\,}O\ensuremath{\,\textsc{iv]}}}

\def\kms{{\rm\,km\,s^{-1}}}

\def\s{\ifmmode \widetilde \else \~\fi}
\def\={\overline}

\def\spose#1{\hbox to 0pt{#1\hss}}

\def\lta{\mathrel{\spose{\lower 3pt\hbox{$\mathchar"218$}}
     \raise 2.0pt\hbox{$\mathchar"13C$}}}
\def\gta{\mathrel{\spose{\lower 3pt\hbox{$\mathchar"218$}}
     \raise 2.0pt\hbox{$\mathchar"13E$}}}
\def\simlt{\lower.5ex\hbox{\ltsima}}
\def\gtsima{$\; \buildrel > \over \sim \;$}
\def\simgt{\lower.5ex\hbox{\gtsima}}

\slugcomment{Accepted by The Astrophysical Journal Supplement}

\shorttitle{The Pan-STARRS1 distant $\lowercase{z}>5.6$ quasar survey}
\shortauthors{Ba\~{n}ados et al.}
\begin{document}

\title{The Pan-STARRS1 distant $\lowercase{z}>5.6$ quasar survey:\\ more than 100 quasars within the first Gyr of the universe}

\author{
E.~Ba\~{n}ados\altaffilmark{1,2,$\dagger$}, 
B.~P.~Venemans\altaffilmark{1},
R.~Decarli\altaffilmark{1},
E.~P.~Farina\altaffilmark{1},
C.~Mazzucchelli\altaffilmark{1},
F.~Walter\altaffilmark{1},
X.~Fan\altaffilmark{3},
D.~Stern\altaffilmark{4},
E.~Schlafly\altaffilmark{1,5,$\ddagger$},
K.C.~Chambers\altaffilmark{6},
H-W.~Rix\altaffilmark{1},
L.~Jiang\altaffilmark{7},
I.~McGreer\altaffilmark{3},
R.~Simcoe\altaffilmark{8},
F.~Wang\altaffilmark{3,9},
J.~Yang\altaffilmark{3,9},
E.~Morganson\altaffilmark{10},
G.~De~Rosa\altaffilmark{11},
J.~Greiner\altaffilmark{12},
M.~Balokovi\'{c}\altaffilmark{13},
W. S. Burgett\altaffilmark{6},
T.~Cooper\altaffilmark{8},
P. W. Draper\altaffilmark{14},
H.~Flewelling\altaffilmark{6},
K. W. Hodapp\altaffilmark{6},
H.~D.~Jun\altaffilmark{4},
N. Kaiser\altaffilmark{6},
R.-P. Kudritzki\altaffilmark{6},
E. A. Magnier\altaffilmark{6},
N. Metcalfe\altaffilmark{14},
D.~Miller\altaffilmark{8},
J.-T.~Schindler\altaffilmark{3},
J.~L.~Tonry\altaffilmark{6},
R.~J. Wainscoat\altaffilmark{6},
C.~Waters\altaffilmark{6},
Q.~Yang\altaffilmark{3,9}
}

\altaffiltext{1}{Max Planck Institut f\"ur Astronomie, K\"onigstuhl 17, D-69117, Heidelberg, Germany}
\altaffiltext{2}{The Observatories of the Carnegie Institute of Washington, 813 Santa
Barbara Street, Pasadena, CA 91101, USA}%
\email{ebanados@carnegiescience.edu}
\altaffiltext{3}{Steward Observatory, The University of Arizona, 933 North Cherry Avenue, Tucson, AZ 85721--0065, USA}
\altaffiltext{4}{Jet Propulsion Laboratory, California Institute of Technology, 4800 Oak Grove Drive, Pasadena, CA 91109, USA}
\altaffiltext{5}{Lawrence Berkeley National Laboratory, One Cyclotron Road, Berkeley, CA 94720, USA} 
\altaffiltext{6}{Institute for Astronomy, University of Hawaii, 2680 Woodlawn Drive, Honolulu, HI 96822, USA} 
\altaffiltext{7}{Kavli Institute for Astronomy and Astrophysics, Peking University, Beijing 100871, China} 
\altaffiltext{8}{MIT-Kavli Center for Astrophysics and Space Research, 77 Massachusetts Avenue, Cambridge, MA, 02139, USA}
\altaffiltext{9}{Department of Astronomy, School of Physics, Peking University, Beijing 100871, China} 
\altaffiltext{10}{National Center for Supercomputing Applications, University of Illinois at Urbana-Champaign, 1205 W. Clark Street, Urbana, IL 61801, USA} 
\altaffiltext{11}{Space Telescope Science Institute, 3700 San Martin Drive, Baltimore, MD 21218, USA} 
\altaffiltext{12}{Max-Planck-Institut f\"ur extraterrestrische Physik, Giessenbachstrasse 1, 85748 Garching, Germany}
\altaffiltext{13}{Cahill Center for Astronomy and Astrophysics, California Institute of Technology, Pasadena, CA 91125, USA}
\altaffiltext{14}{Department of Physics, Durham University, South Road, Durham DH1 3LE, UK} 
\altaffiltext{$\dagger$}{Carnegie-Princeton Fellow}
\altaffiltext{$\ddagger$}{Hubble Fellow}

\begin{abstract}
Luminous quasars at $z>5.6$ can be studied in detail with the current generation of telescopes and provide us with unique information on the first gigayear of the universe. 
Thus far these studies have been statistically limited by the number of quasars known at these redshifts. 
Such quasars are rare and therefore wide-field surveys are required to identify them and multiwavelength data are needed to separate them efficiently from their main contaminants, the far
more numerous cool dwarfs. In this paper, we update and extend the selection for $z\sim 6$ quasars presented in Ba\~nados et al. (2014) using the Pan-STARRS1 (PS1) survey. 
We present the PS1 distant quasar sample, which currently consists of 124 quasars in the redshift range $ 5.6 \lesssim z  \lesssim 6.7$ that satisfy our selection criteria. Seventy-seven of these quasars have been discovered with PS1, and 63 of them are newly identified in this paper. We present composite spectra of the PS1 distant quasar sample.  This sample spans a factor of $\sim 20$ in luminosity and shows a variety of emission line properties.  The number of quasars at $z>5.6$ presented in this work almost double the quasars previously known at these redshifts,  marking a transition phase from studies of individual sources to statistical studies of the high-redshift quasar population, which was impossible with earlier, smaller samples.
\end{abstract}


\vfil
\eject
\clearpage

\section{INTRODUCTION}
\label{sec:intro}

Quasars---accreting supermassive black holes in the center of massive galaxies---have fascinated astronomers since their discovery over fifty years ago \citep{schmidt1963}. The extreme luminosity of quasars 
makes them light beacons that literally illuminate our knowledge of the early universe. Within only a few years of their initial discovery, quasars with redshifts as high as $z\sim 2$ (i.e., when the universe was about one quarter of its current age) were being identified \citep[e.g.,][]{schmidt1965,arp1967}. This allowed astronomers to study  objects at distances that at the time were unconceivable, expanding our view of the universe. 

About fifteen years ago the first quasars at $z>5.6$ (i.e., within the first gigayear of the universe) were discovered \citep{fan2000}. 
By 2011 the number of quasars at $z>5.6$ reached to 60, with most of the contributions coming from large surveys such as the SDSS \citep[e.g.,][]{fan2006a,jiang2008}, CFHQS \citep[e.g.,][]{willott2007,willott2010a}, and UKIDSS \citep[e.g.,][]{venemans2007a,mortlock2011}. Over the last three years a second wave of quasar discoveries has started due to new optical and near-infrared large sky surveys such as VIKING \citep{venemans2013}, ATLAS \citep{carnall2015}, DES \citep{reed2015}, HSC \citep{matsuoka2016}, and Pan-STARRS1 \citep[PS1; e.g.,][]{banados2014}.

Quasars within the first gigayear of the universe place strong constraints on black hole formation models \citep{volonteri2012a} and are fundamental probes of the final phases of cosmological reionization \citep[see][for recent reviews]{beckerG2015b,mortlock2015}. However, the conclusions provided are still fairly weak, due to the low number of quasar sightlines studied so far.

Over the last years, we have been searching for high-redshift quasars in the PS1 survey \citep{kaiser2002,kaiser2010}, which has imaged the whole sky above a declination of $-30\degr$ for about four years in five filters (\gps, \rps, \ips, \zps, \yps). Our efforts have resulted in 14  published quasars at $z>5.6$  \citep{morganson2012,banados2014,banados2015a,venemans2015}, one of which is among the brightest quasars known in the early universe in both UV and \cii\ luminosities \citep{banados2015b}. In this work we update our selection criteria and present 63 new quasars. The 77 PS1-discovered $z>5.6$ quasars almost double the number of quasars previously known at these redshifts, which give us the opportunity to perform the initial characterizations of the high-redshift quasar population as a whole.

The quasars presented in this paper were selected from the first and second internal releases of the stacked PS1 data (PV1 and PV2 in the internal naming convention, respectively). 
At the time of writing this article the PV3 and final version of the PS1 catalog was made available to us. Therefore, throughout the paper we quote 
the PV3 PS1 magnitudes corrected for Galactic extinction\footnote{Note that the PS1 magnitudes presented in \cite{banados2014,banados2015a,venemans2015} were 
not corrected for Galactic extinction.} \citep{schlafly2011}.

 This article is organized as follows. In Section \ref{sec:ps1-selection}, we introduce our updated 
color selection criteria for $i$-dropout $5.7 \lesssim z \lesssim 6.5$ quasars  from
the PS1 stacked catalog and summarize our candidate selection procedures. The imaging and spectroscopic follow-up observations are 
presented in Section \ref{sec:ps1-followup}. 
In Section \ref{sec:ps1-discoveries}, we show the spectra 
of the 77 PS1-discovered quasars at $5.7 \lesssim z \lesssim 6.7$ (63 newly discovered in this paper) and discuss some individual objects. 
In Section \ref{sec:ps1-sample}, we introduce the PS1 distant quasar sample, which currently consists of 124 quasars at $z>5.6$ that were 
discovered by PS1 or that satisfy the selection criteria presented in this work or in \cite{venemans2015}. We create composite spectra from the PS1 distant quasar sample in Section \ref{sec:composite} and
revisit the discussion of how typical weak-line quasars are at $z\sim 6$ in Section \ref{sec:wlq}.  Finally, we summarize our results in Section \ref{sec:summary}.

The International Astronomical Union naming convention for non-transient objects discovered using the \PS\ survey is 
``PSO~JRRR.rrrr+DD.dddd''  where RRR.rrrr and +DD.dddd are the right ascension and declination in decimal degrees (J2000), respectively. 
For PS1-discovered quasars we will use abbreviated names of the form ``PRRR+DD'', while quasars discovered by other surveys will be named 
as ``Jhhmm+ddmm''. Table \ref{tab:qsos-info} in the Appendix \ref{append:qsolist} lists the coordinates and redshifts of all 173 $z>5.6$ quasars 
 known to date\footnote{A machine readable format can be obtained from the online journal. An updated version can be obtained upon request from the authors.}.

All magnitudes are given in the AB system. When referring to limiting 
magnitudes (\mbox{mag}\ensuremath{_{\rm P1,lim}}) throughout the text,  these 
 correspond to $3\sigma$-limiting magnitudes. We use a flat $\Lambda$CDM cosmology 
with $H_0 = 67.7 \,\mbox{km\,s}^{-1}$\,Mpc$^{-1}$, $\Omega_M = 0.307$, and $\Omega_\Lambda = 0.693$ \citep[][]{planck2015XIII}.

\section{Candidate Selection}

\label{sec:ps1-selection}

\begin{figure}[ht]
\centering
\includegraphics[scale=0.62]{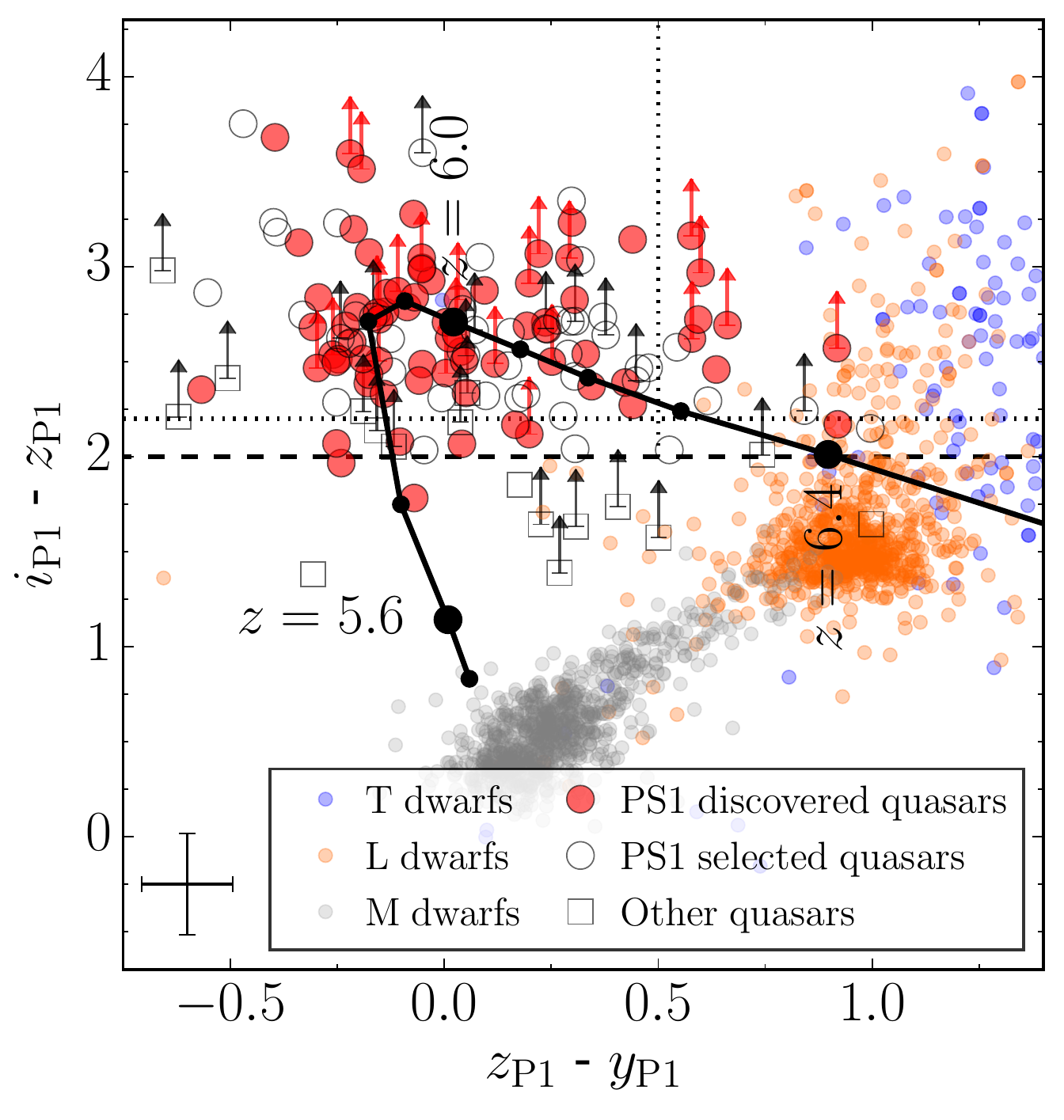}
\caption[PS1 color selection criteria.]{
 Color--color diagram showing the criteria used 
to select quasar candidates (dashed and dot-dashed lines, see text).
 The thick black line shows the expected color of
the PS1 composite quasar spectrum created in Section \ref{sec:composite} redshifted from $z=5.5$ to $z=6.5$ in steps of $\Delta z = 0.1$. 
The L/T dwarfs that have a PS1 counterpart are shown with orange/blue small circles. 
A subsample of the M dwarfs from \cite{west2011} are represented by gray small circles. Upper limits for brown dwarfs are not plotted to enhance the 
clarity of the figure. 
Red filled circles are quasars discovered with \PS\ in this work, \cite{morganson2012}, and \cite{banados2014,banados2015a}. Empty circles represent known quasars discovered by other surveys
that satisfy the \PS\ selection criteria presented in this paper. Empty squares show known quasars discovered by other surveys that 
do not comply with the PS1 selection criteria. A representative error bar is shown in the bottom left corner.
}
\label{fig:ps1-selection}
 \end{figure}

The main contaminants of $z\sim 6-7$ quasar searches are brown dwarfs, especially late M-, L-, and T-dwarfs, which can have similar optical colors to quasars but are much more abundant. 
Figure \ref{fig:ps1-selection} shows the expected location of the composite quasar spectrum created in Section \ref{sec:composite} in the   $\ips - \zps$ vs. $\zps - \yps$  color space as its redshift is increased from $z=5.5$ to $z=6.5$.  In order to visualize the PS1 colors of brown dwarfs, we cross-matched the M-dwarf catalog of \cite{west2011} to the PS1 PV3 catalog. We also cross-matched our compilation of 1827 L and T dwarfs\footnote{Based on the list compiled by \cite{mace2014}, with additions from 
\cite{lodieu2014}, \cite{marocco2015}, and \cite{best2015}.
} 
with the PS1 stacked catalog, taking the closest match within a $ 2\arcsec$ radius. There are 986 matches (734 and 252 L and T dwarfs, respectively) with
measurements in the \ips, \zps, and \yps\ bands, and with a signal-to-noise ratio (S/N) $> 5$ in the \zps\ and \yps\ bands.
The PS1 colors of brown dwarfs are represented by small circles in Figure \ref{fig:ps1-selection}.

 In this paper we update and extend the $i$-dropout selection ($z\sim 6$ quasars) discussed in detail in \cite{banados2014}. The $z$-dropout selection ($z\sim 7$ quasars) is discussed in \cite{venemans2015} and an update will be given by Mazzucchelli et al. (in prep). 

  Given the rarity and faintness of these high-redshift quasars plus the numerous foreground objects that can have similar PS1 colors, finding these quasars efficiently poses a big challenge. For that reason, we followed several steps in order to clean up our candidates list. 
 In short, we selected initial high-redshift quasar candidates from the PS1 PV1 or PV2 database (Section \ref{sec:ps1-catalog}). This was followed by forced photometry at the position of each candidate in both their stacked and single-epoch images to corroborate the catalog colors and remove artifacts (this process removes about $80\%$ of the initial candidates; see Sections 2.2 and 2.3 in \citealt{banados2014}). We then matched the candidates list to public infrared surveys to eliminate or prioritize candidates using the extra information provided by these surveys (Section \ref{sec:ir_surveys}). 
 Before following-up the candidates, we visually inspected their PS1 stacked and single-epoch images (and infrared images when available) to ensure that they were real astrophysical objects. We then obtained optical and near-infrared follow-up photometry to remove lower redshift interlopers (Section \ref{sec:ps1-photrometry-followup}) and then finally acquired spectra of the remaining candidates (Section \ref{sec:ps1-spectroscopy-followup}).

 \subsection{The \PS\ catalog}
\label{sec:ps1-catalog}

\begin{figure}[ht]
\centering
\includegraphics[scale=0.65]{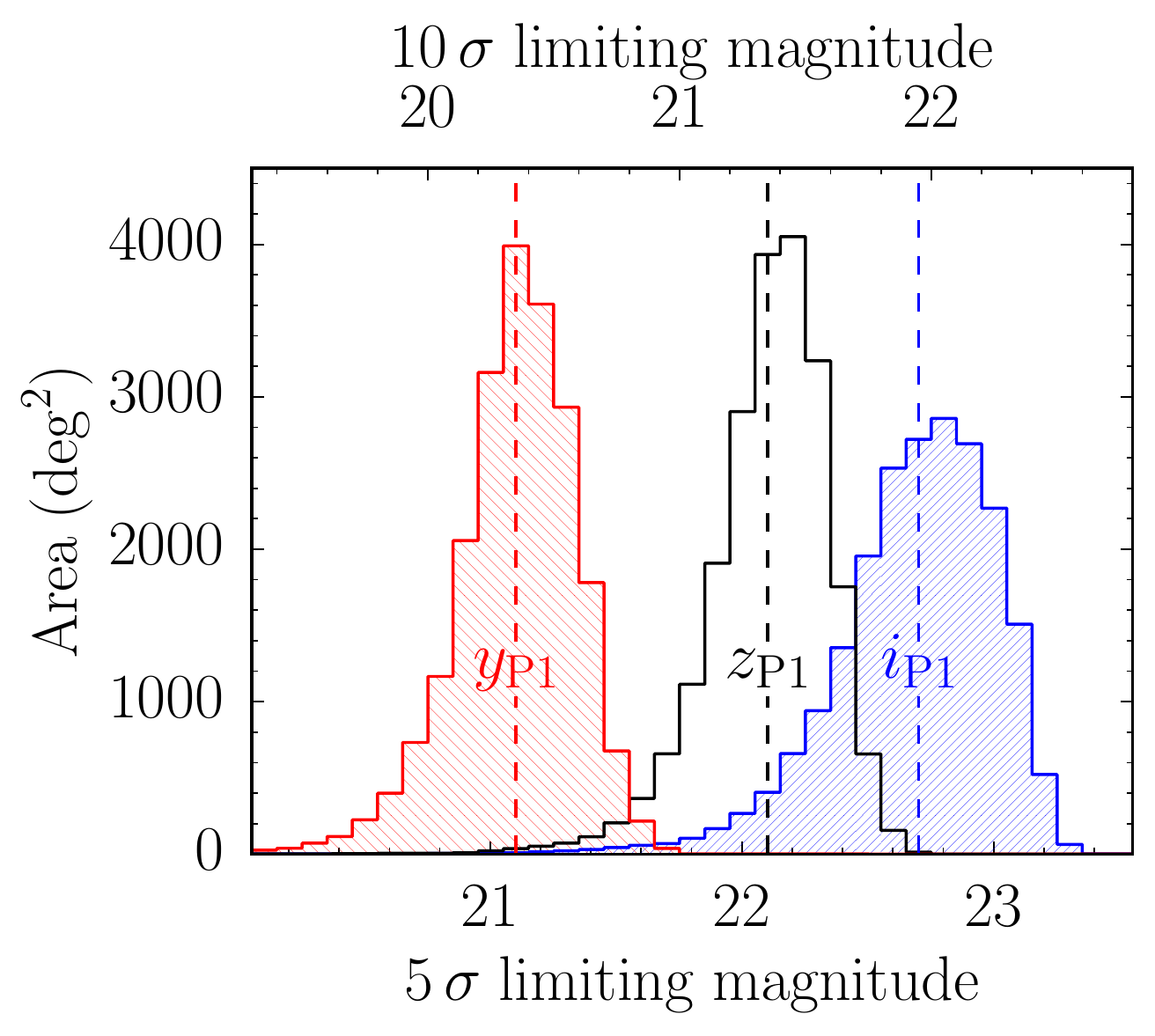}
\caption{
PS1 PV2 5$\sigma$ extinction corrected limiting magnitude distributions in our quasar search area for the main bands used in our quasar selection: $\ips$, $\zps$, and $\yps$ (see Section \ref{sec:ps1-catalog}). The vertical dashed lines show the median magnitudes of the survey. For reference the top axis shows the respective $10\sigma$ limiting magnitudes.
}
\label{fig:limmag}
 \end{figure}

The selection presented here is based on the PS1 PV1 and PV2 catalogs. Figure \ref{fig:limmag} shows the PV2 5$\sigma$ (and $10\sigma$) extinction corrected $\ips$, $\zps$, and $\yps$ limiting magnitude distributions in our search area. 
The $5\sigma$ median limiting 
magnitudes are (\gps, \rps, \ips, \zps, \yps) = (23.2, 23.0, 22.7, 22.1, 21.1). 
As in our previous works, we excluded candidates in the Milky Way plane ($\left| b \right| < 20\degr$) and M31 ($7\degr< \,$R.A.$ <14\degr$; $37\degr< \,$Decl.$ <43\degr$), this yields a survey effective area of $2.05\,\pi$ steradians.  However, this time we followed up a few bright quasar candidates in the Galactic plane taking advantage of the \cite{schlegel1998} dust maps, requiring a reddening of $E(B-V)<0.3$. This resulted in the discovery of two $z\sim 6$ quasars with $\left | b \right|<20\degr$, a region that had not been explored by other high-redshift quasar surveys (see Section \ref{sec:individual_notes}). While we did not exclude any area around M33 (R.A.\,$\sim23\degr$ and Decl.\,$\sim30\degr$), candidates in that region   were more critically inspected since the number of candidates in that area is larger than in other regions of the sky.   

We also excluded those measurements for which the Image Processing Pipeline  \citep[IPP;][]{magnier2006,magnier2007} flagged the result as suspicious
(see Table 6 in \citealt{banados2014}).  
Furthermore, we required that more than  $85\%$ of the expected point-spread function (PSF)-weighted flux in the \ips,\zps, and \yps\ bands was 
located in valid pixels (i.e., that the PS1 catalog entry has 
{\tt PSF\_QF} $> 0.85$). 

\begin{figure}[ht]
\centering
\includegraphics[scale=0.7]{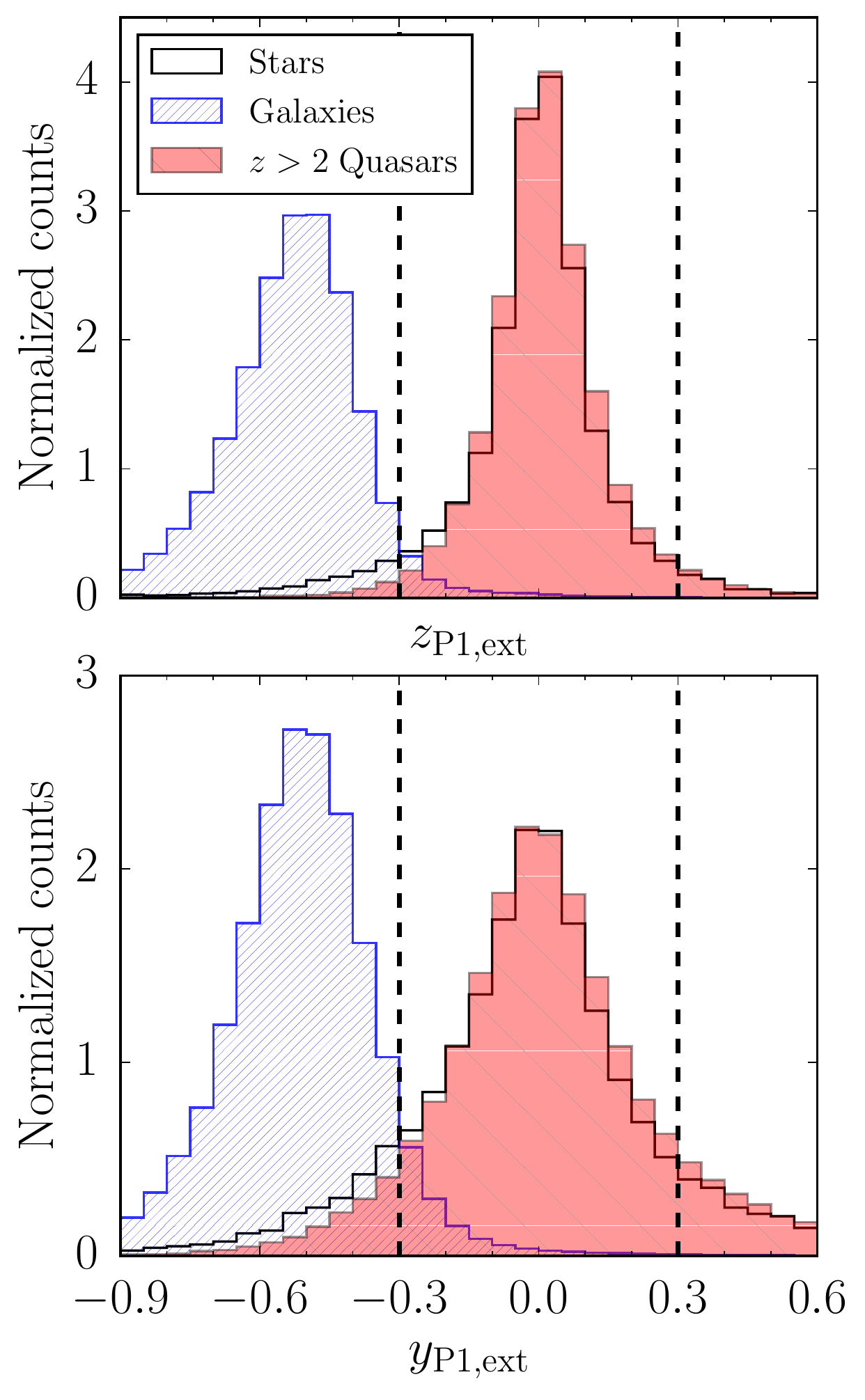}
\caption{
Aperture minus PSF magnitudes (\mbox{mag}\ensuremath{_{\rm P1,ext}}) for the \zps\ (top) and \yps\ (bottom) bands for different sources as indicated in the legend. 
The quasars are taken from the SDSS-DR10 quasar catalog \citep{paris2014}, while SDSS spectroscopic stars and galaxies are retrieved from the region $160\degr< \,$R.A.$ <220\degr$; $0\degr< \,$Decl.$ <20\degr$. All sources in this figure have S/N(\zps)$\;>10$, S/N(\yps)$\;>5$,  $\zps>19$, and $\yps>19$. The counts are normalized so that the integrals of the histograms sum to 1 (bin width $=0.05$). As part of our quasar candidate selection, we require ($-0.3<\zext < 0.3$) \textit{or} \;  ($-0.3< \yext <0.3$) (dashed vertical lines).
}
\label{fig:fext}
 \end{figure}

We use the difference between the aperture and  PSF magnitudes (\mbox{mag}\ensuremath{_{\rm P1,ext}})  as a proxy to remove extended sources. As a test, we obtained the PS1 information for all $z>2$ quasars from the SDSS-DR10 quasar catalog \citep{paris2014}, and a sample of spectroscopic stars and galaxies from SDSS \citep[DR12;][]{alam2015}. Figure \ref{fig:fext} shows the \zext\ and \yext\ histograms for these sources, with the requirement that S/N(\zps)$\;>10$, S/N(\yps)$\;>5$, $\zps>19$, and $\yps>19$.  The PSF magnitudes for galaxies are systematically fainter than the aperture magnitudes as expected. 
Since we are interested in point sources, we require our candidates to satisfy ($-0.3<\zext < 0.3$) \textit{or}   ($-0.3< \yext <0.3$). With these criteria we exclude 92\% of the galaxies while recovering 93\% and 97\% of stars and quasars, respectively.

\subsubsection{\textit{i}-dropout search ($5.7 \lesssim z \lesssim 6.5$)}
\label{sec:idrops1}

In \cite{banados2014} we searched for quasars in the upper left  region of Figure \ref{fig:ps1-selection} (i.e., $\zps - \yps < 0.5$ and $\ips - \zps > 2.2$). This is the PS1 color space region where quasars in the redshift range  $5.7 \lesssim z \lesssim 6.2$
are best differentiated from brown dwarfs.
This is the selection on which we initially focused and therefore the one with the most comprehensive follow-up. In the present work we extend the search to $\zps - \yps > 0.5$. This allows us to search for quasars in the redshift range $6.2 \lesssim z \lesssim 6.5$ but admittedly in a region highly contaminated, especially by L and T dwarfs (see the upper right region in Figure \ref{fig:ps1-selection}). Another difference with respect to the selection in \cite{banados2014} is that now our initial PS1 color criteria are based on the \textit{dereddened} magnitudes. This allows us to relax our criteria to $\ips -\zps > 2.0$ but we  still prioritize the reddest objects (i.e., $\ips - \zps > 2.2$, horizontal dotted line in Figure \ref{fig:ps1-selection}) for follow-up. Furthermore, in contrast to \cite{banados2014}, we do not impose any flux restriction (besides the implicit S/N cuts); i.e., a source can be unrestrictedly bright. 
The selection criteria can be summarized as follows:

\begin{subequations}
\begin{eqnarray}
((\mbox{S/N}(\ips) \geq 3 )\; \mbox{AND} \, (\ips - \zps > 2.0)) \; \mbox{OR} & \nonumber \\
 (\ipsl - \zps > 2.0) & \label{eq:i1-izcolor} \\
\mbox{S/N}(\gps) < 3  & \label{eq:i1-gcolor} 
\end{eqnarray}
\end{subequations}

\noindent Additionally, candidates with $\zps - \yps < 0.5$ are required to comply with the following:

\begin{subequations}
\begin{eqnarray}
\mbox{S/N}(\zps) >  10& \label{eq:i1-zsn} \\
\mbox{S/N}(\yps) > 5 & \label{eq:i1-ysn} \\
\mbox{S/N}(\rps) < 3 \; \mbox{OR}  \; (\rps - \zps > 2.2) & \label{eq:i1-rzcolor}  
\end{eqnarray}
\end{subequations}

\noindent While the requirements for candidates with $\zps - \yps \geq 0.5$ are:

 \begin{subequations}
\begin{eqnarray}
\mbox{S/N}(\zps) >  7& \label{eq:i2-zsn} \\
\mbox{S/N}(\yps) > 7 & \label{eq:i2-ysn} \\
\mbox{S/N}(\rps) < 3  \label{eq:i2-rzcolor}  
\end{eqnarray}
\end{subequations}

\subsection{Public infrared surveys}
\label{sec:ir_surveys}
We matched our sources with several public infrared surveys to extend and verify the photometry of the quasar candidates.
The extra information was used to either remove foreground interlopers or to prioritize the subsequent follow-up. 

\begin{figure*}[ht]
\centering
\includegraphics[scale=0.5]{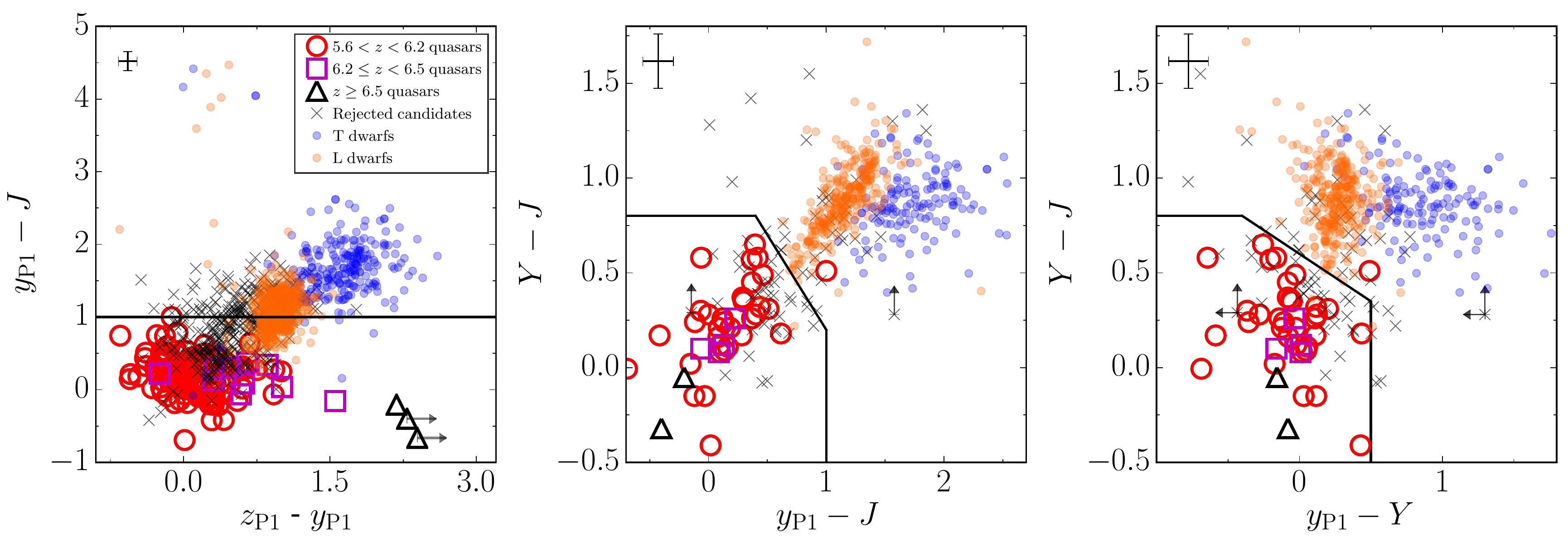}
\caption[Quasar selection criteria when the $Y$ and/or $J$ bands are available.]{
Selection criteria when the $Y$ and/or $J$ bands are available (black solid lines).
Empty red circles, magenta squares, and black triangles represent the colors of known quasars at 
$5.6<z<6.2$, $6.2 \leq z<6.5$, and $z\geq6.5$, respectively.
L and T dwarfs are shown with orange and blue small circles, respectively. Candidates rejected by follow-up photometry are shown as gray crosses. 
Upper limits for brown dwarfs are not displayed to enhance the 
clarity of the figure. Representative error bars are shown in the upper left corner of each panel. The magnitudes in this figure are not dereddened. 
\textit{Left}: $\zps - \yps$ vs. $\yps - J$ color--color diagram.
\textit{Middle}: $\yps -J$ vs. $Y - J$ color--color diagram.
\textit{Right}: $\yps - Y$ vs. $Y - J$ color--color diagram.
}
\label{fig:yj-selection}
 \end{figure*}

 \textit{2MASS:} The PS1 candidates were matched within $3\arcsec$  with the Two Micron All Sky Survey \citep[2MASS;][]{skrutskie2006}. 
 This is a shallow all-sky survey in the $J$, $H$, and $K$ bands, with nominal $5\sigma$ limiting AB magnitudes of $17.5$, $17.3$, and $16.9$, respectively. Nevertheless, given its large areal coverage, 
 it is ideal to eliminate bright foreground interlopers  and even find extremely bright high-redshift quasars \citep[e.g., ][]{wu2015}.
 In order to remove bright cool dwarfs, we required our candidates to be undetected in 2MASS or to have $\yps - J<1$ (see Figure \ref{fig:yj-selection}).

 \textit{UKIDSS:} We matched our objects with the near-infrared data from the UKIDSS survey \citep{lawrence2007} using a $2\arcsec$ matching radius.
 The UKIDSS Large Area Survey provides $Y$, $J$, $H$, and $K$ imaging over $\sim$4000\,deg$^2$, with nominal 5$\sigma$ AB limiting magnitudes of 21.1, 20.9, 20.2, and 20.3, respectively.  
 We kept the candidates that had $Y-J<0.8$, $\yps -J < 1$ , $\yps - Y < 0.5$, and $Y-J < -(y-J) + 1.2$ (see Figure \ref{fig:yj-selection}).
 
  \textit{VHS:} We cross-matched our candidates to the $J$-band catalog of the first data release of the VISTA Hemisphere Survey \citep[VHS;][]{mcmahon2013}. This release covers  $\sim$8000\,deg$^2$ 
 to a $5\sigma$ limiting AB magnitude of $J =  21.1$.  We applied the same color criteria as for our 2MASS matched list.

\begin{figure*}[ht]
\centering
\includegraphics[scale=0.55]{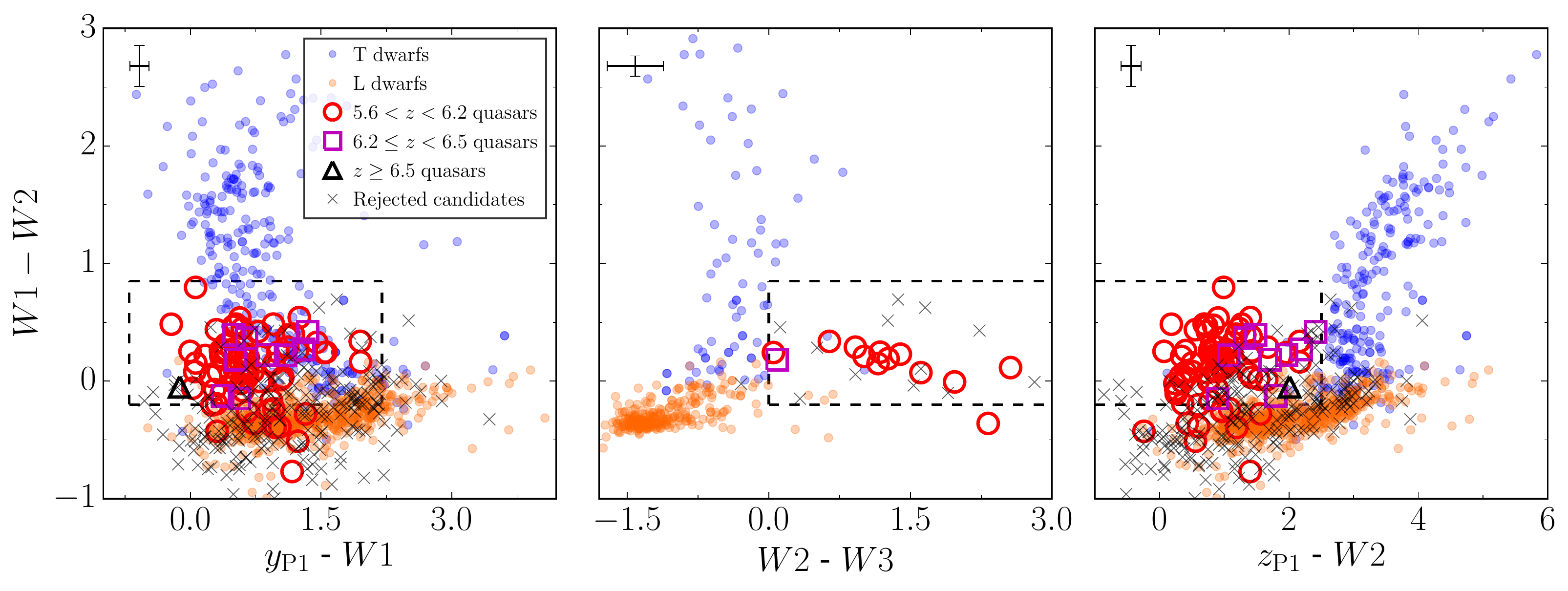}
\caption[Prioritization criteria for candidates detected in \textit{WISE} with $\mbox{S/N}>3$.]{
Prioritization criteria for candidates detected in \textit{WISE} with $\mbox{S/N}>3$ (black dashed lines).
Empty red circles, magenta squares, and black triangles represent the colors of known quasars at 
$5.6<z<6.2$, $6.2 \leq z<6.5$, and $z\geq6.5$, respectively.
L and T dwarfs are shown with orange and blue small circles, respectively. Candidates rejected by follow-up photometry are shown as gray crosses. 
Upper limits for brown dwarfs are not displayed to enhance the 
clarity of the figure. Representative error bars are shown in the upper left corner of each panel. The magnitudes in this figure are not dereddened. 
\textit{Left}: $\yps - W1$ vs. $W1 - W2$ color--color diagram.
\textit{Middle}: $W2 - W3$ vs. $W1 - W2$ color--color diagram.
\textit{Right}: $\zps -W2$ vs. $W1 - W2$ color--color diagram. This criteria is not used for $z$-dropout candidates (c.f., \citealt{venemans2015}).
}
\label{fig:wise-crit}
 \end{figure*}

 \textit{WISE:} \textit{WISE} \citep{wright2010} surveyed the entire mid-infrared sky in four bands centered at 3.4, 4.6, 12, and 22\,$\mu$m 
 (hereafter $W1$, $W2$, $W3$, and $W4$). In regions that are not confusion limited,  the nominal $5\sigma$ limiting AB magnitudes of the ALLWISE catalog\footnote{\url{http://wise2.ipac.caltech.edu/docs/release/allwise}} are $W1=19.6$, $W2=19.3$, $W3=16.7$, and $W4=14.6$.  
 Even though more than half of the known $z>6$ quasars are detected in \textit{WISE}, a selection 
 of high-redshift quasars purely based on \textit{WISE} is extremely 
 difficult owing to their \textit{WISE} colors being hard to distinguish from 
 active galactic nuclei (AGNs) and star-forming galaxies at lower  
 redshifts \citep{blain2013}. Nevertheless, the combination of \textit{WISE} and optical surveys is a powerful tool to remove or avoid a large
 fraction of the main foreground contaminants of optical surveys, i.e.,  cool dwarfs (see Figure \ref{fig:wise-crit}).
 We cross-matched our quasar candidates with the ALLWISE catalog within $3\arcsec$ (but see Appendix \ref{append:photometry}). At this stage, we only used the \textit{WISE} information
  to prioritize candidates for follow-up observations. Objects with $\mbox{S/N}>3$ in $W1$ and $W2$ were assigned a higher priority if their 
 colors fulfilled the additional criteria: 
 
 \begin{subequations}
\begin{eqnarray}
-0.2 < W1 - W2 < 0.85 & \nonumber  \\
 -0.7 < \yps - W1 < 2.2 \nonumber
\end{eqnarray}
\end{subequations}

\noindent For the few objects with $\mbox{S/N}>3$ in $W3$, a higher priority was assigned if
\begin{equation}
 W2 - W3 > 0  \nonumber \\
\end{equation}

 \noindent Recently, in part motivated by the quasar selection criteria presented in \cite{carnall2015}, we added an additional prioritization for 
$i$-dropout candidates:

\begin{equation}
\zps- W2 < 2.5.  \nonumber \\
\end{equation}

The dashed lines in Figure \ref{fig:wise-crit} show our prioritization criteria.
Objects  undetected in the ALLWISE catalog or with $\mbox{S/N}<3$ in $W1$ or $W2$ were assigned an intermediate priority, while the remaining 
candidates were given a low priority.  Note that we have not rejected candidates because of their \textit{WISE} colors 
and we have even discovered quasars that do not fulfill our prioritization criteria.

\section{Follow-up observations}
 \label{sec:ps1-followup}
\subsection{Photometry}
\label{sec:ps1-photrometry-followup}
Since many of our candidates have PS1 magnitudes close to our S/N cuts, we obtained deep optical follow-up imaging to corroborate the PS1 colors and eliminate objects that were scattered into our color selection. Additionally, we obtained deep near-infrared imaging, which provides essential information to separate efficiently cool dwarfs---our main contaminants---from high-redshift quasars (Figure \ref{fig:yj-selection}).
The photometric follow-up observations were carried out over different observing runs and different instruments.
We obtained optical and near-infrared images with the MPG 2.2\,m/GROND \citep{greiner2008}, New Technology Telescope (NTT)/EFOSC2 \citep{buzzoni1984}, NTT/SofI \citep{moorwood1998}, 
 the Calar Alto (CAHA) 3.5\,m/Omega2000 \citep{bizenberger1998,bailer-jones2000}, the CAHA 2.2\,m/CAFOS\footnote{\url{www.caha.es/CAHA/Instruments/CAFOS/index.html}}, the MMT/SWIRC \citep{brown2008}, and the du~Pont/Retrocam\footnote{\url{www.lco.cl/telescopes-information/irenee-du-pont/instruments/website/retrocam}}; see Table \ref{tab:imaging_obs} for details of the 
 observations and filters used.

We reduced the data and obtained the zero points following standard procedures (e.g., see Section 2.6 in \citealt{banados2014}). The near-infrared data taken with the 2.5\,m du~Pont telescope 
were reduced by collaborators from the Carnegie Supernova Project,
with dark subtraction, flat fielding, and bad-pixel masked final combination
as detailed in \cite{hamuy2006}. For completeness, we provide below the color conversions used to calibrate our follow-up imaging:

\begin{subequations}
\begin{eqnarray*}
\ggrond = &  \gps + 0.332 \times (\gps - \rps) + 0.055 \\
\rgrond = & \rps + 0.044 \times (\rps - \ips) - 0.001\\
\igrond = & \ips - 0.089 \times (\rps - \ips) + 0.001  \label{eq:PS_iGROND}\\
\zgrond = &\zps - 0.214 \times (\zps - \yps)   \label{eq:PS_zGROND}\\
\Jgrond = &\Jtmass - 0.012 \times (\Jtmass - \Htmass) + 0.004 \label{eq:PS_JGROND}\\
\Hgrond = &\Htmass + 0.030 \times (\Htmass - \Ktmass) + 0.009 \label{eq:PS_HGROND}\\
\intt = &\ips - 0.149 \times (\ips - \zps) - 0.001  \label{eq:PS_iNTT}\\
\zntt = &\zps - 0.265 \times (\zps - \yps)  \label{eq:PS_zNTT}\\
\zotk = & \zps - 0.245 \times (\zps - \yps) \\
\Yotk = & \yps - 0.413 \times (\zps - \yps) +0.012\\
\Jotk = & \Jtmass + 0.093  \times (\Jtmass - \Htmass)\\
\Icafos = & \ips - 0.098 \times (\ips - \zps)
\end{eqnarray*}
\end{subequations}

\noindent where \Jtmass, \Htmass, and \Ktmass\ are 2MASS magnitudes in the AB system. 
\Jntt, $J_{\rm SWIRC}$, \jdp, and \Hotk\ are calibrated against 2MASS.

 Candidates were considered foreground interlopers if they had $Y-J>0.8$, $\yps -J > 1$ , $\yps - Y>0.5$,  or $Y-J > -(y-J) + 1.2$ 
 (see Section \ref{sec:ir_surveys} and Figure \ref{fig:yj-selection}). 
 
\begin{figure*}[ht]
\centering
\includegraphics[scale=0.5]{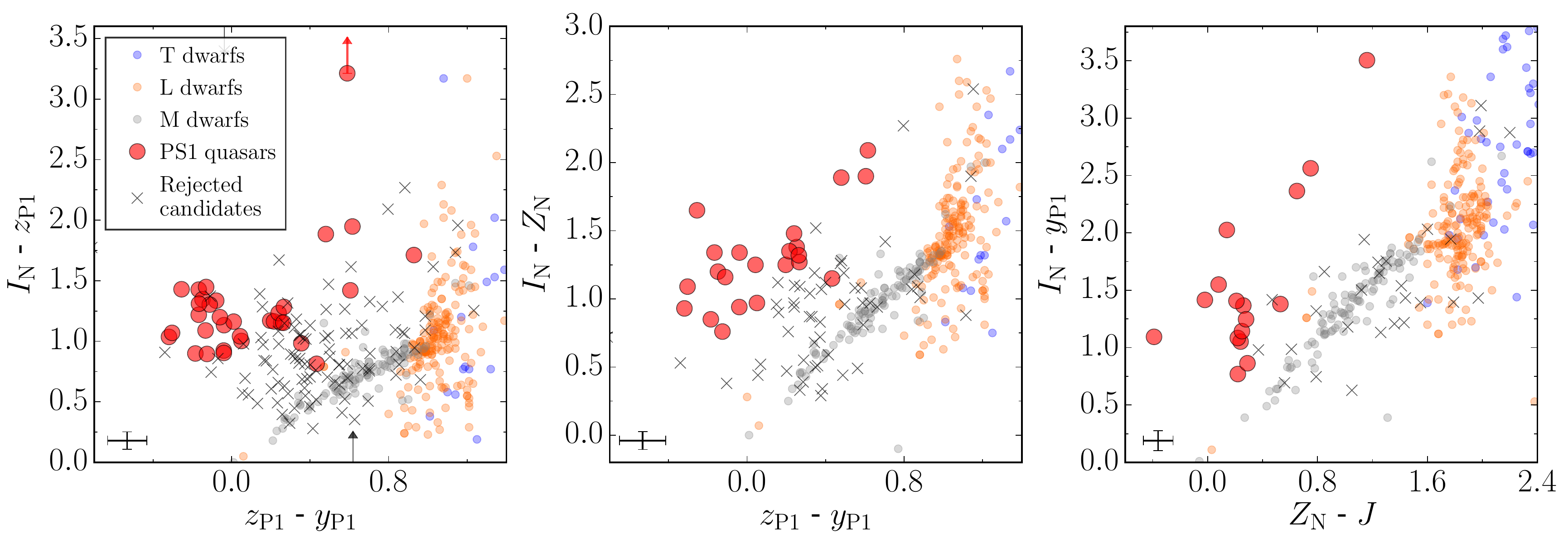}
\caption{
Diagrams used to select candidates for follow-up spectroscopy based on their NTT imaging follow-up.  The brown dwarf colors in this figure are synthetic colors obtained from spectra taken from the SpeX Prism Spectral Libraries\footnote{\url{http://pono.ucsd.edu/~adam/browndwarfs/spexprism/}}. Candidates rejected by follow-up photometry are shown as gray crosses.  Representative error bars are shown in the lower left corner of each panel. The magnitudes in this figure are not dereddened. 
\textit{Left}: $\zps - \yps$ vs. $\intt - \zps$ color--color diagram.
\textit{Middle}: $\zps - \yps$ vs. $\intt - \zntt$ color--color diagram.
\textit{Right}: $\zntt - J$ vs. $\intt - \yps$ color--color diagram.
}
\label{fig:ntt-crit}
 \end{figure*}

 The NTT/EFOSC2 filters \intt\ (\#705) and \zntt\ (\#623) are significantly different from the $\ips$ and $\zps$ filters. For candidates observed with \intt, \zntt, and/or $J$ bands, we used the color-color diagrams in Figure~\ref{fig:ntt-crit} to select targets for spectroscopic follow-up. 

Candidates rejected by at least one of our photometric follow-up criteria are shown as crosses in Figures \ref{fig:yj-selection}, \ref{fig:wise-crit}, and \ref{fig:ntt-crit}. 
Tables \ref{tab:follow-idrops1} and  \ref{tab:follow-idrops2} present the imaging follow-up of the PS1-discovered quasars 
selected with the criteria of Sections \ref{sec:idrops1} with $\zps -\yps <0.5$ and $\zps -\yps \geq 0.5$, respectively. Table \ref{tab:follow-extended} lists
the imaging follow-up of three PS1-discovered quasars that were selected using a more relaxed selection criteria in terms of their 
optical colors and two that were found at low Galactic latitudes ($\left| b \right| < 20\degr$). These five objects are discussed further in Section \ref{sec:individual_notes}.

\begin{deluxetable*}{lcccc}
\tablecolumns{4}
\tablewidth{0pc}
\tablecaption{Imaging observations of quasar candidates \label{tab:imaging_obs}}
 \tablehead{
 \colhead{Date} &   \colhead{Telescope/Instrument}   &  \colhead{Filters} & \colhead{Exposure Time} 
}
\startdata
 2012 May 21--24       & MPG 2.2\,m/GROND        &  \ggrond,\rgrond,\igrond,\zgrond,\Jgrond,\Hgrond,\Kgrond        & 460--1440\,s        \\%
 2013 Jan 14--18       & MPG 2.2\,m/GROND        &  \ggrond,\rgrond,\igrond,\zgrond,\Jgrond,\Hgrond,\Kgrond        & 460--1440\,s        \\%
 2013 Jan 26           & CAHA 3.5\,m/Omega2000   &   \zotk,\Yotk,\Jotk  & 900\,s \\ 
 2013 Mar 13--16       & NTT/EFOSC2              &  \intt, \zntt        & 300\,s \\ 
 2013 Mar 23--29       & CAHA 3.5\,m/Omega2000   &   \zotk,\Yotk,\Jotk  & 300\,s \\
 2013 Apr 16--18       & CAHA 2.2\,m/CAFOS       &   \Icafos            & 1000\,s \\ 
 2013 Apr 26--27       & CAHA 3.5\,m/Omega2000   &   \zotk,\Yotk,\Jotk  & 300--600\,s \\
 2013 Aug 18--19       & CAHA 3.5\,m/Omega2000   &   \zotk,\Yotk,\Jotk  & 300--600\,s \\
 2013 Sep 7--10        & MPG 2.2\,m/GROND        &  \ggrond,\rgrond,\igrond,\zgrond,\Jgrond,\Hgrond,\Kgrond        & 460--1440\,s        \\%
 2013 Sep 27--Oct 1    & NTT/EFOSC2              &  \intt, \zntt        & 600\,s \\ 
 2013 Oct 16--21       & CAHA 3.5\,m/Omega2000   &   \zotk,\Yotk,\Jotk,\Hotk  & 300\,s \\ 
 2013 Nov 9--12        & CAHA 2.2\,m/CAFOS       &   \Icafos            & 1250\,s \\ 
 2013 Nov 15--17       & CAHA 3.5\,m/Omega2000   &   \zotk,\Yotk,\Jotk  & 300\,s \\
 2013 Dec 14--15       & CAHA 2.2\,m/CAFOS       &   \Icafos            & 1500\,s \\ 
 2014 Jan 24--Feb 5    & MPG 2.2\,m/GROND        &  \ggrond,\rgrond,\igrond,\zgrond,\Jgrond,\Hgrond,\Kgrond        & 460--1440\,s        \\%
 2014 Mar 2--6         & NTT/EFOSC2              &  \intt, \zntt        & 600\,s \\ 
 2014 Mar 2 and 5      & NTT/SofI                &  \Jntt               & 300\,s \\ 
 2014 Mar 16--19       & CAHA 3.5\,m/Omega2000   &   \zotk,\Yotk,\Jotk  & 300\,s \\ 
 2014 Apr 22--24       & CAHA 2.2\,m/CAFOS       &   \Icafos            & 2400\,s \\ 
 2014 May 9            & CAHA 3.5\,m/Omega2000   &   \zotk,\Yotk,\Jotk  & 300\,s \\ 
 2014 Jul 23--27       & NTT/EFOSC2              &  \intt, \zntt        & 600\,s \\ 
 2014 Jul 25           & NTT/SofI                &  \Jntt               & 600\,s \\ 
 2014 Aug 7 and 11--13 & CAHA 3.5\,m/Omega2000   &   \Yotk,\Jotk        & 600\,s \\ 
 2014 Aug 22--24       & CAHA 2.2\,m/CAFOS       &   \Icafos            & 1800\,s \\ 
 2014 Sep 12 and 14    & CAHA 3.5\,m/Omega2000   &   \Yotk,\Jotk        & 600\,s \\ 
 2014 Sep 16--17       & CAHA 2.2\,m/CAFOS       &   \Icafos            & 1800\,s \\ 
 2014 Sep 17--25       & MPG 2.2\,m/GROND        &  \ggrond,\rgrond,\igrond,\zgrond,\Jgrond,\Hgrond,\Kgrond        & 460--1440\,s        \\%
 2014 Dec 13--21       & MPG 2.2\,m/GROND        &  \ggrond,\rgrond,\igrond,\zgrond,\Jgrond,\Hgrond,\Kgrond        & 460--1440\,s        \\%
 2015 Feb 7 and 28     & CAHA 3.5\,m/Omega2000   &  \zotk,\Yotk,\Jotk    & 600--900\,s \\ 
 2015 Feb 19--23       & NTT/EFOSC2              &  \intt,\zntt        & 600\,s \\ 
 2015 Feb 22            & NTT/SofI                &  \Jntt               & 300\,s \\ 
 2015 Mar 1 and 11-12   & CAHA 3.5\,m/Omega2000   &   \zotk,\Yotk,\Jotk,\Hotk  & 600--900\,s \\ 
 2015 Apr 13--15       & CAHA 2.2\,m/CAFOS       &   \Icafos            & 1800\,s \\ 
2015 May 18--28      & MPG 2.2\,m/GROND        &  \ggrond,\rgrond,\igrond,\zgrond,\Jgrond,\Hgrond,\Kgrond        & 460--1440\,s        \\%
2015 Jun 8             & MMT/SWIRC              &  $J_{\rm SWIRC}$              & 300\,s \\ 
 2015 Jul 21--23       & NTT/EFOSC2              &  \intt,\zntt        & 600\,s \\ 
 2015 Jul 20 and 23        & NTT/SofI                &  \Jntt               & 300\,s \\
 2015 Aug 7--13       & MPG 2.2\,m/GROND        &  \ggrond,\rgrond,\igrond,\zgrond,\Jgrond,\Hgrond,\Kgrond        & 460--1440\,s        \\%
2015 Sep 15--20        & CAHA 2.2\,m/CAFOS       &   \Icafos            & 1800\,s \\ 
 2015 Sep 28--Oct 1       & CAHA 3.5\,m/Omega2000   &   \zotk,\Jotk  & 300--600\,s \\
2015 Nov 4--8      & MPG 2.2\,m/GROND        &  \ggrond,\rgrond,\igrond,\zgrond,\Jgrond,\Hgrond,\Kgrond        & 460--1440\,s        \\%
 2016 Jan 12       & CAHA 2.2\,m/CAFOS       &   \Icafos            & 3360\,s \\ 
 2016 Jan 30--Feb 1--2       &  NTT/SofI             &  \Jntt      & 300\,s \\ 
 2016 Jan 31--Feb 2--3       &      NTT/EFOSC2            &   \intt,\zntt          & 300--900\,s \\
 2016 March 30--31       &      du Pont/RetroCam            &  \ydp,\jdp          & 220--1200\,s 
\enddata
\end{deluxetable*}

\subsection{Spectroscopy}
\label{sec:ps1-spectroscopy-followup}

We have spectroscopically followed-up candidates that satisfied the selection from the previous sections. This spectroscopic campaign was carried out using several instruments at different telescopes: EFOSC2
at the NTT telescope in La Silla, the FOcal Reducer/low dispersion Spectrograph 2 \citep[FORS2;][]{appenzeller1992}
at the Very Large Telescope (VLT), 
the Folded-port InfraRed Echellette \citep[FIRE;][]{simcoe2008, simcoe2013} spectrometer and the Low Dispersion Survey Spectrograph (LDSS3) at the Baade and Clay telescopes at Las Campanas Observatory,
the  Low Resolution Imaging Spectrometer \citep[LRIS;][]{oke1995} at the Keck I 10\,m telescope on Mauna Kea, 
the Double Spectrograph  \citep[DBSP;][]{oke1982} on the 200-inch (5\,m) Hale telescope at Palomar Observatory (P200), 
the Red Channel Spectrograph \citep{schmidt1989} on the 6.5\,m MMT Telescope, the Cassegrain TWIN spectrograph at the 3.5\,m Calar Alto telescope (CAHA3.5\,m), 
the Multi-Object Double Spectrograph  \citep[MODS;][]{pogge2010} and the LUCI spectrograph \citep[][]{seifert2003} at 
the Large Binocular Telescope (LBT).

The details of the spectroscopic observations of the PS1-discovered quasars are shown in Table \ref{tab:spectroscopy}. There were 11 candidates for which the spectroscopy revealed a non-quasar interloper, their photometric information is presented in Appendix \ref{append:rejected}. 
The spectra were reduced using standard routines including bias subtraction, flat fielding, sky subtraction, and wavelength calibration using exposures of He, HgCd and Ne arc lamps.

\section{77 new quasars at $z>5.6$}
\label{sec:ps1-discoveries}
We have discovered 77 quasars at $5.6<z<6.7$ mining the PS1 database, out of which 63 are new discoveries presented in this paper. In \cite{morganson2012} and \cite{banados2014,banados2015a} we presented our first 11 $i$-dropout ($z\sim 6$) quasars, while our first 3 $z$-dropout ($z>6.5$) quasars were introduced in \cite{venemans2015}. For completeness we include these quasars in the tables and figures of this section.  The spectra of all PS1 quasars discovered to date are shown in Figure \ref{fig:qso-spectra}. The $i$-dropout spectra are scaled to match their dereddened \zps\  magnitude while the $z$-dropouts are scaled to their near-infrared magnitudes (see \citealt{venemans2015}). 

The names, redshifts, and coordinates of these newly discovered quasars are listed in Table \ref{tab:qsos-info} and their dereddened PS1 PV3 magnitudes and corresponding $E(B-V)$ values  are 
given in Table \ref{tab:qsos-ps1wiseinfo}.

\subsection{Redshifts}
\label{sec:redshifts}

Estimating accurate quasar systemic redshifts from broad emission lines is challenging. At $z>5.6$, it is especially difficult as the most prominent emission lines in optical spectra are Ly$\alpha$, which is highly affected by absorption at these redshifts, and high-ionization lines such as \civ\  and 
\siivpoiv,  which are known to be poor estimators of systemic redshifts \citep{richards2002a,derosa2014,shen2016}. Ideally, we would prefer to measure the redshifts from atomic or molecular emission lines such as \cii\ and CO, which provide host galaxy redshifts as accurate as $\Delta\, z <0.002$ \cite[e.g.,][]{wang2010,banados2015b}. Alternatively, for quasars at $6.0\lesssim z \lesssim 7.4$, the low ionization line \mgii\ can be observed in the near-infrared $K$-band. This line is thought to be a more accurate tracer of the systemic redshift compared to the high-ionization lines, although at $z\sim 6$ significant shifts of $480\pm 630\, \kms$ are found between the redshift of the \mgii\ line and that of atomic or molecular lines (see discussion in \citealt{venemans2016}). In this paper, we report redshifts estimated from fittings to the \cii, CO, or \mgii\ lines when available.

However, for the vast majority of the quasars presented here only optical spectra are available, several of which show only weak emission lines. 
We estimate the redshifts for these quasars following a template fitting approach. We perform a $\chi^2$ minimization relative to two quasar templates. One is the composite $z\sim 6$ SDSS quasar spectra from  \cite{fan2006a} and the other is the template from \cite{selsing2016}. The latter was chosen over other typical quasar templates because it is created only from bright quasars, which have comparable luminosities to the $z\sim 6$ quasars discovered in this work. The redshifts derived from these two templates generally agree well but sometimes differences of up to 0.05 are observed. In these cases, the fits are visually inspected to assess which one is the best match. Conservatively, for these cases we report redshifts with significance up to the second decimal only.

\subsection{Rest-frame 1450\AA\ magnitudes}
\label{sec:m1450}

Another important quantity is the magnitude at rest-frame 1450\,\AA\ ($m_{1450}$), which plays a key role in the estimation of the quasar luminosity function. At $z>5.6$ the rest-frame 1450\,\AA\ is shifted to observed wavelengths $>9570\,$\AA. Thus, $m_{1450}$ is hard to estimate for $z>5.6$ quasars, especially when only optical spectra are available. 

The main challenge is to determine the continuum which is mostly prevented by the low rest-frame wavelength coverage of optical spectra ($ \lambda_{\rm rest} < 1500$\,\AA) and often also low S/N. In the literature, several methods are used to estimate these quantities, including interpolation and fitting a power law of the form $f_\nu = C \times \nu^{\alpha_{\nu}}$ to regions of the continuum that are generally uncontaminated by
emission lines. In the latter case,  a good fit of the continuum is typically not possible unless the power law index $\alpha_{\nu}$ is fixed. A range of $\alpha_{\nu}$ indices are used by different authors, the most common being $\alpha_{\nu}=-0.5$, which is consistent with the average quasar UV continuum slope found by \cite{vandenberk2001}. Recently, \cite{selsing2016} found that $\alpha_{\nu}=-0.3$ is a good fit for luminous quasars without significant host contamination, as is expected for our luminous $z>5.6$ quasars. Furthermore, some of the $m_{1450}$ values quoted in the literature were estimated with outdated redshifts when much more accurate values are currently available. This fact and given that the literature values for $m_{1450}$ are not calculated in a consistent manner could eventually have repercussions on other fundamental measurements such as the quasar luminosity function.

In order to determine $m_{1450}$ in a consistent way and circumvent the fact that most of our spectra have a limited wavelength coverage and some of them have poor S/N, we adopt the following approach. We assume a power law continuum slope $\alpha_{\nu}=-0.3$ \citep{selsing2016}. For quasars in the redshift range $5.6<z<6.3$ we calculate $m_{1450}$ by extrapolating the $\yps$ magnitude ($\lambda_{\rm eff} = 9627.7\,$\AA). In this redshift range, the extrapolation to rest-frame $1450$\,\AA\ is of the order of $\sim100$\,\AA\ or less. In some cases the \yps\ band will be contaminated by the \siivpoiv\ emission line which has a typical (rest-frame) equivalent width of 8\,\AA\ \citep{vandenberk2001}. Given the width of the \yps\ band of $\sim 800$\,\AA\, \citep{tonry2012}, \siivpoiv\ would contribute less than $\sim 10\%$ of the flux in the broad band. At $z\geq 6.3$ contributions from Ly$\alpha$ and$ \nv$\ might start being significant for the flux measured in \yps. Therefore, at these redshifts we extrapolate $m_{1450}$ from their $J$-band magnitude ($\lambda_{\rm eff} = 12444.0\,$\AA). The $J$ band is in a region clean of strong emission lines up to $z\sim 7$ where is contaminated by \civ, which has a typical (rest-frame) equivalent width of $\sim 24$\,\AA. Even though \civ\ is brighter than \siivpoiv, the $J$ band is also wider than the \yps\ band ($\sim 1500$\,\AA) and thus the contamination should be of comparable order. The extrapolations from the $J$-band magnitude to $m_{1450}$ are of the order of $100-300$\,\AA. We also computed the rest-frame magnitude assuming a slope of $\alpha_{\nu}=-0.5$, and find that the absolute differences are negligible (mean: 0.02, standard deviation: 0.02, median: 0.01). This is due to the effective wavelengths of our chosen filters which minimize the extrapolation.

We followed this approach for all published $z>5.6$ quasars with information in the \yps\ or $J$ bands. For comparison, the $m_{1450}$ literature values compiled by \cite{calura2014} differ from ours by $-0.03 \pm 0.31$ mag. 
The observed and absolute magnitudes at rest-frame 1450\,\AA\ are listed in Table \ref{tab:qsos-info}.

\begin{figure*}[h!]
\centering
\includegraphics[scale=0.8]{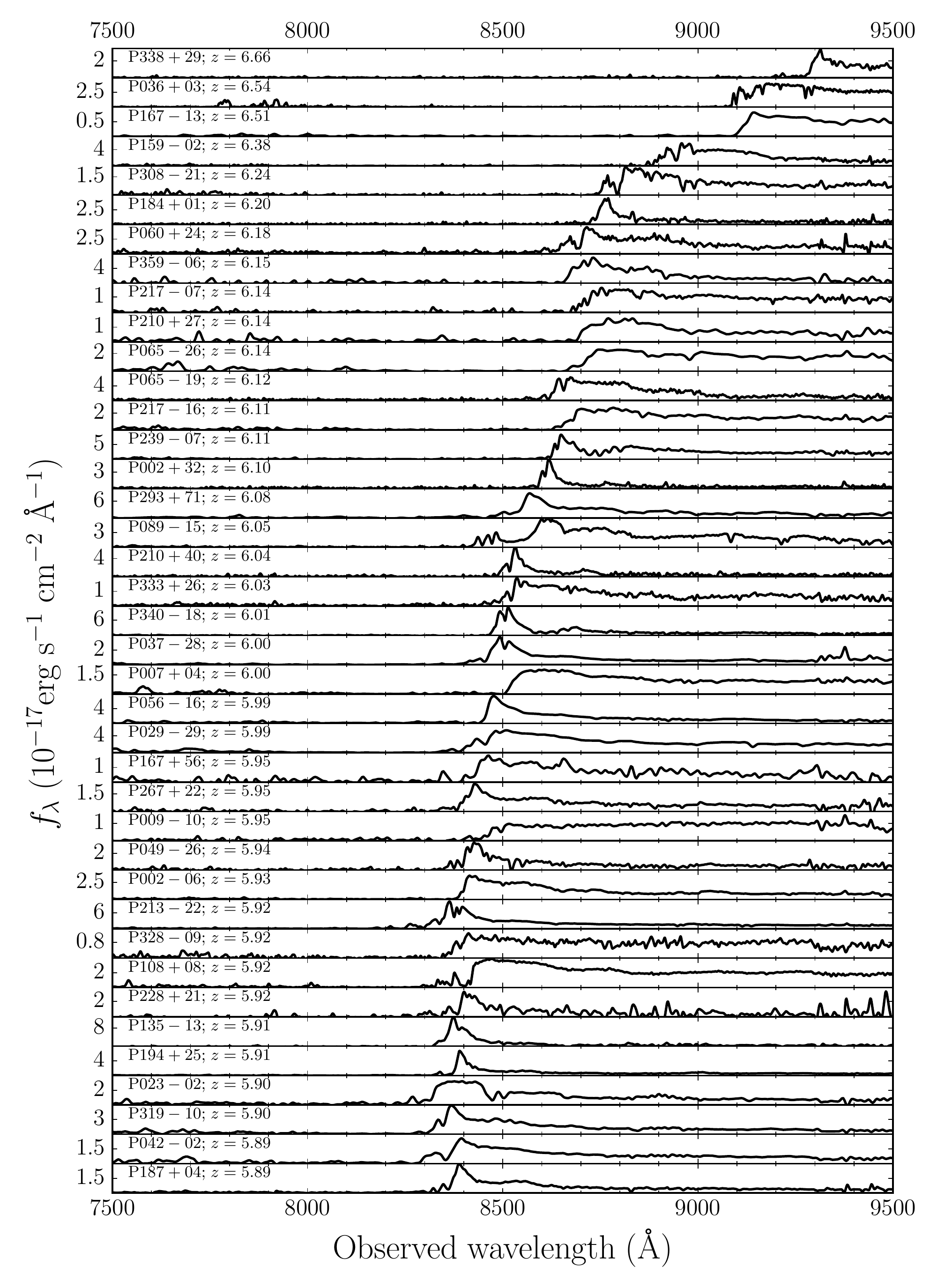}
 \caption{Spectra of the 77 Pan-STARRS1 discovered quasars at $z\geq 5.6$. 
 Sorted by decreasing redshift.
 \label{fig:qso-spectra}}
 \end{figure*}
 
\begin{figure*}[h!]
\centering
\includegraphics[scale=0.8]{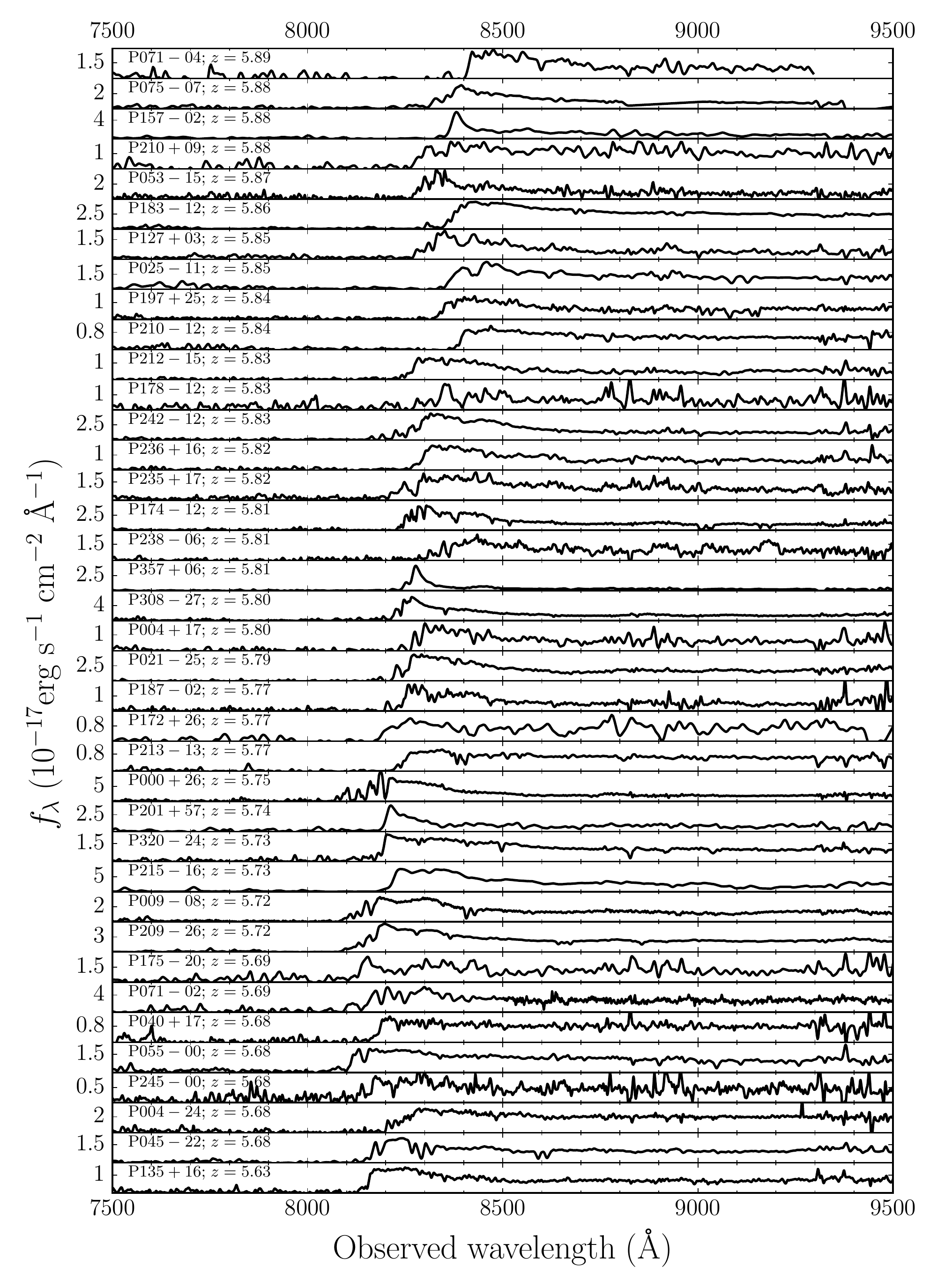}\\

{\textbf{Figure~\ref{fig:qso-spectra}}: Continuation}
 \end{figure*}

\subsection{Notes on selected objects}
\label{sec:individual_notes}

In this section we discuss some of the new quasars, including five PS1 quasars that were not selected using the criteria of Section \ref{sec:ps1-selection}.
The objects are sorted by R.A.

\subsubsection{PSO~J000.3401+26.8358 ($z=5.75$)}
This quasar has a color $\ips -\zps =1.97$ and therefore just below our selection criteria presented in Section \ref{sec:ps1-selection}.
However, it was targeted for follow-up because it was relatively bright ($\zps=19.28$) and it did satisfy all the other criteria including good 
\textit{WISE} and $\yps - J$ colors (see Section \ref{sec:ir_surveys} and Figures \ref{fig:wise-crit} and \ref{fig:yj-selection}). 

\subsubsection{PSO~J055.4244--00.8035 ($z=5.68$)}
This quasar was discovered by the selection criteria presented in \cite{banados2015a}. 
That selection was based on the PS1 PV1 catalog and further required a radio-counterpart in the 
Faint Images of the Radio Sky at Twenty cm survey \citep[FIRST;][]{becker1995}, which allowed us to relax the requirements on
optical colors. 
This is among the radio loudest quasar known at $z>5.5$ (see Figure 2 in \citealt{banados2015a}).
While the PV1 $\ips - \zps$ color for this quasar was a lower limit 
 of $\ips - \zps >1.3$, its PV3  $\ips - \zps$ color is a detection of $\ips - \zps = 2.08$. Therefore, with the deeper PV3 \ips\ band this quasar does satisfy the selection criteria of Section \ref{sec:ps1-selection}.

 \subsubsection{PSO~J089.9394--15.5833  ($z=6.05$)}
This quasar is located at a Galactic latitude of $b=-18.31\degr$, in a region with $E(B-V)=0.30$. This is the first $z>5.6$ quasar discovered within $\left| b \right| < 20\degr$ of the Milky Way plane. In addition, with $\yps-J=1.0$ and $Y-J=0.51$ it does not satisfy the criteria of Figure \ref{fig:yj-selection}, although its $Y$-band follow-up was taken after the discovery of this quasar (see Tables \ref{tab:imaging_obs} and \ref{tab:follow-extended}).

   \subsubsection{PSO~J108.4429+08.9257 ($z=5.92$)}
Similar to P089--15, this quasar is located  within the Galactic plane with $b=8.99\degr$ and $E(B-V)=0.09$. Unlike P089--15 though, P108+08 satisfies our color selection. The discovery of both P089--15 and P108+08 opens up a new area for quasar searches: regions closer to the Milky Way plane but with low extinction values ($E(B-V)<0.3$). ß

\subsubsection{PSO~J135.3860+16.2518 ($z=5.63$)}
Similar to  P055--00, this quasar was selected by the PS1/FIRST criteria presented in \cite{banados2015a}.
P135+16 is also among the radio loudest quasars at $z>5.5$ (see also Appendix \ref{append:photometry}). 
Its PV1 and PV3 $\ips -\zps$ colors do not differ much: $\ips-\zps=1.70$ vs. $\ips-\zps=1.78$.  This color is too blue 
for the color cuts typically applied in optical searches for high-redshift quasars  (for instance, the color cuts presented in this work). 
Thus, this quasar would have been 
missed if  not for its strong radio emission.

 \subsubsection{PSO~J245.0636--00.1978 ($z=5.68$)}
This quasar was selected from the PV2 database with a S/N(\yps)$=5$, i.e., just within the limit of our selection criteria (see Eq. \ref{eq:i1-ysn}). However, the PV3 $\yps$ band of this object has S/N(\yps)$=4.5$ and we would have therefore missed this quasar even though it satisfies every other criteria.

 \subsubsection{PSO~J210.8296+09.0475 ($z=5.88$)}
  This quasar was independently discovered by \cite{jiang2015} 
 and they reported a slightly lower redshift ($z=5.86$), but consistent within the uncertainties.

\section{The PS1 distant quasar sample}
\label{sec:ps1-sample}
At the time of writing (2016 March) there are 173 known quasars at $z>5.6$. We provide their names, coordinates, and redshifts in 
Table \ref{tab:qsos-info}; and their PS1 PV3, $J$-band, and \textit{WISE} magnitudes in Table \ref{tab:qsos-ps1wiseinfo}.
The PS1 PV3 catalog has information for about 81\% of these quasars, with at least 5$\sigma$ detections in the \zps\ or \yps\ bands.
The selection criteria presented in Section \ref{sec:ps1-selection} and \cite{venemans2015} recover 119 of these quasars plus five PS1-discovered quasars that were 
selected by extended criteria (see Section \ref{sec:individual_notes}). Thus, the PS1 distant quasar sample currently consists of 124 
quasars at $z>5.6$, encompassing more than 70\% of the quasars known at these redshifts.

\begin{figure*}[ht]
\centering
\includegraphics[scale=0.60]{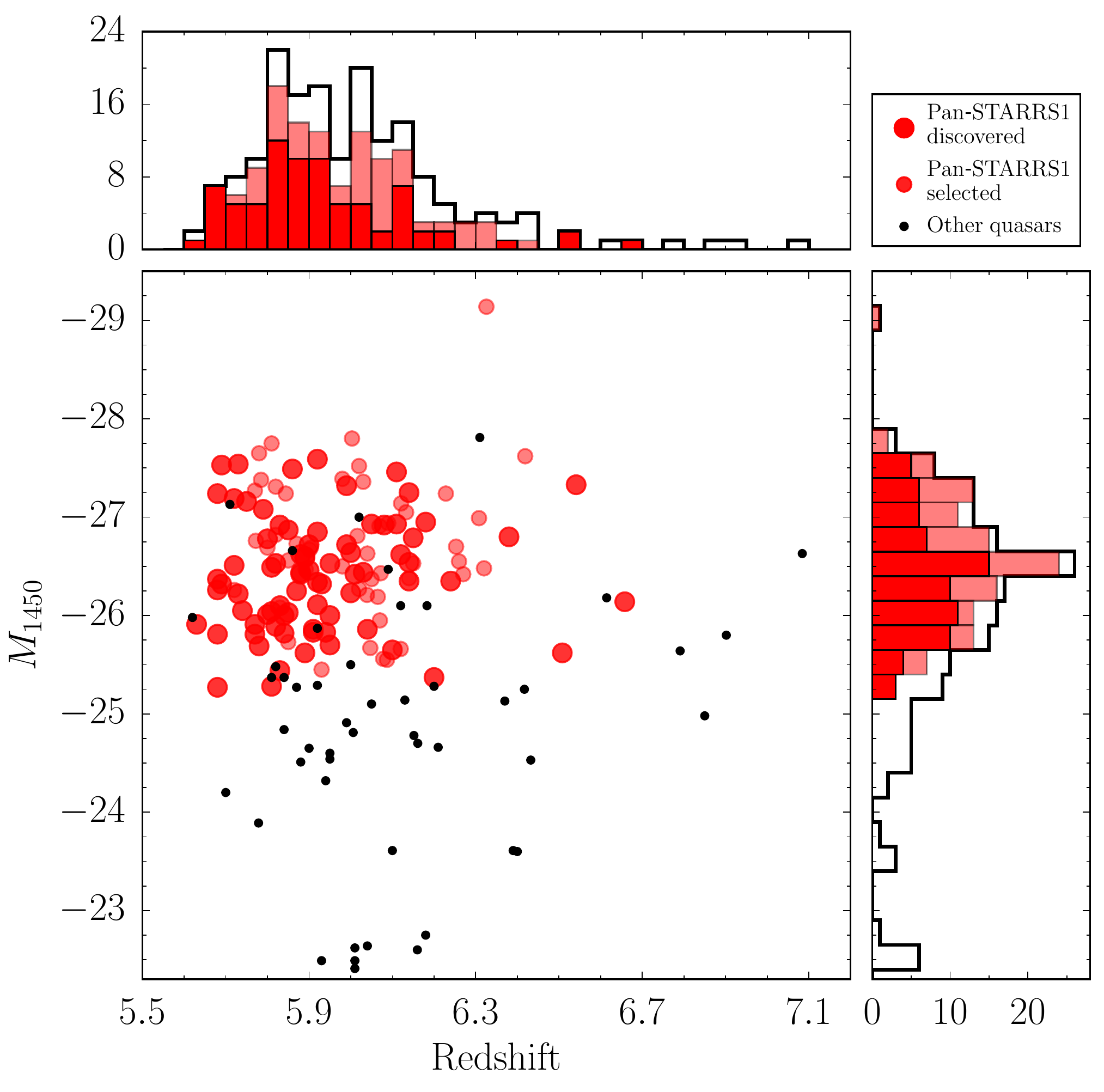}
\caption{
Redshift and absolute UV magnitude ($M_{1450}$) distribution of all the known $z>5.6$ quasars as of 2016 March.
}
\label{fig:z-uv}
 \end{figure*}

Figure \ref{fig:z-uv} shows the redshift and UV luminosity distribution of all known quasars at $z>5.6$, 
highlighting the PS1 distant quasar sample in red. 

\begin{figure*}[ht]
\centering
\includegraphics[scale=0.75]{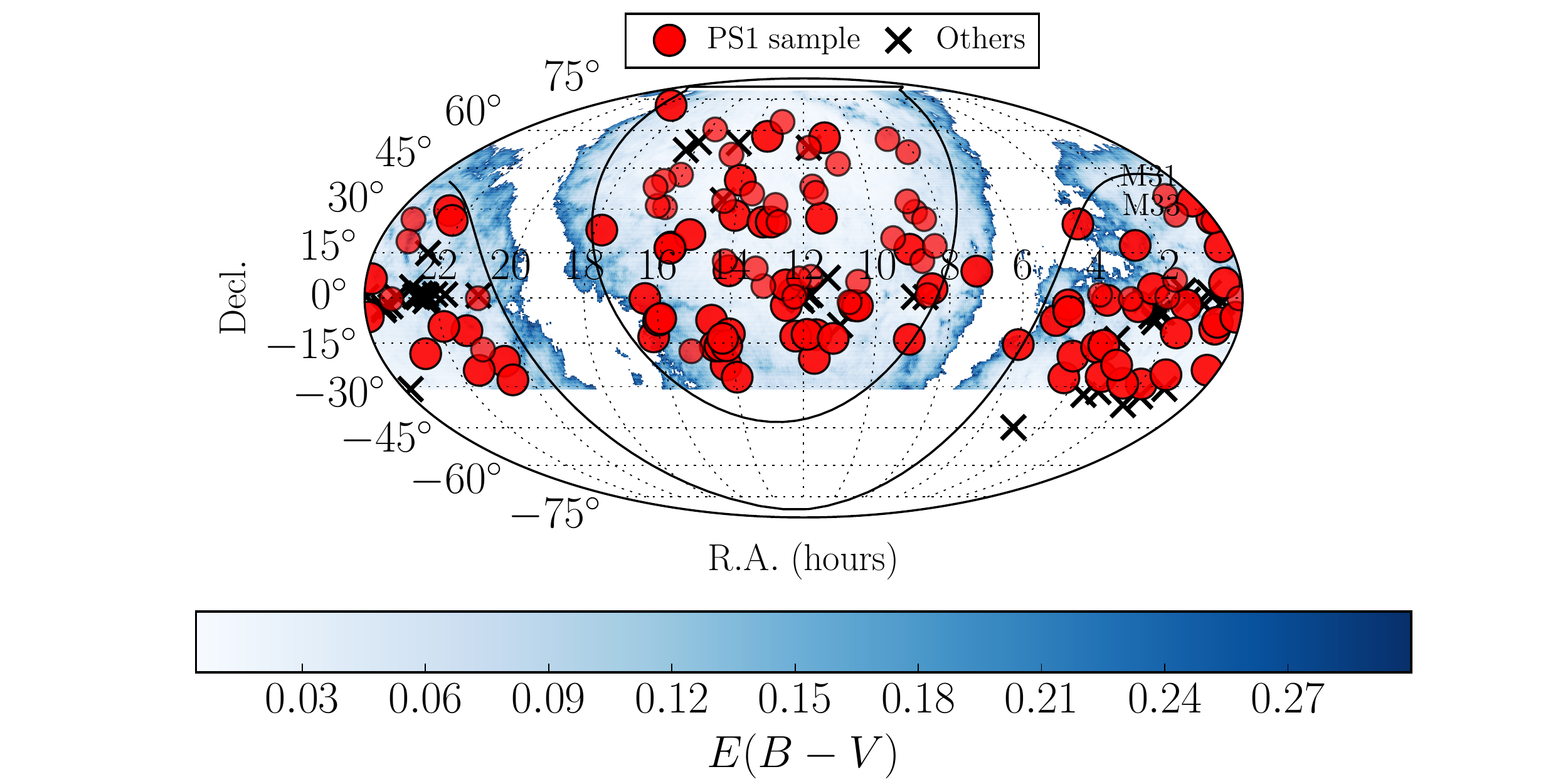}
\caption{
Sky distribution of all known $z>5.6$ quasars. Red circles represent the PS1 distant quasar sample. Larger symbols are PS1 discoveries. Black crosses are quasars that do not belong to the PS1 distant quasar sample. The solid lines show the border of the Milky Way plane traditionally avoided  by high-redshift quasar surveys ($| b | < 20\degr$). The location of M31 and M33 are also indicated. The color map shows the $E(B-V)$ reddening map in the PS1 footprint (Decl.\,$>-30\degr$) from \cite{schlegel1998} where $E(B-V)< 0.3$. Note that we have discovered two quasars with $| b | < 20\degr$ in regions with low extinction.
}
\label{fig:sky-distribution}
 \end{figure*}

The sky distribution of all quasars at $z>5.6$ is presented in Figure \ref{fig:sky-distribution}. We see that a large fraction of the new PS1 discoveries are located in the southern sky. There are seven known quasars at Decl. $<-30\degr$, which are not recovered by our search since they fall outside of the PS1 footprint. The seven quasars at Decl. $<-30\degr$ were discovered by  
\cite{carnall2015},  \cite{reed2015}, and \cite{venemans2013,venemans2015b} using the VST ATLAS, DES, and VIKING and KiDS surveys, respectively.

There are 13 other known quasars that even though they have information in the PS1 database, they do not satisfy our S/N requests or their $\ips$ images are too shallow to satisfy our color cuts (SDSS J0129--0035, CFHQS~J0136+0226, CFHQS J0316--1340, CFHQS J1059--0906, VIK J1148+0056, CFHQS J1429+5447, SDSS J2053+0047, CFHQS J2229+1457, SDSS J2307+0031, CFHQS J2318--0246,   CFHQS J2329--0301, SDSS J2356+0023). We do not select SDSS J1621+5155 ($z=5.71$) given its  $\ips -\zps=1.85$ color, although it does satisfy our $\yps-J$ requirement and all of 
our \textit{WISE} prioritization criteria (see Figures \ref{fig:yj-selection} and \ref{fig:wise-crit}). ULAS~J1120+0641 ($z=7.08$) is a special case as it 
does not appear in the PS1 PV3 catalog. Forced photometry on the PV3 \yps\ image yields a $4.3\,\sigma$ detection with $\yps=21.31\pm 0.25$.  This is significantly fainter than what we expected from its published photometry. 
 This would not be the first time that a possible flux decrement has been suggested for this object. \cite{simpson2014} reported that the HST photometry of this quasar was 14\% (19\%) fainter in the F105W (F125W) filter than expected from its discovery spectrum and photometry \citep{mortlock2011}. We have recently re-observed this quasar in the $Y$ and $J$ bands with the RetroCam instrument at the du Pont telescope in Las Campanas Observatory. The observations were carried out on 2016 March 30 and the total exposure times were 1200\,s. The measured magnitudes are $\ydp=20.36 \pm 0.06$ and $\jdp=20.36 \pm 0.05$. While these magnitudes are fainter than the reported magnitudes for this quasar ($Y=20.26\pm 0.04$, $J=20.16\pm 0.07$; \citealt{barnett2015}), the differences are not significant ($\lesssim 2\sigma$).

\section{PS1 quasar composite spectrum}
\label{sec:composite}

\cite{fan2004} presented the first quasar composite spectrum at $z\sim 6$. This composite consisted of 12 SDSS quasars at $z>5.7$, and showed no clear differences in spectral properties relative to quasars at $z\sim 2$ (see their Figure 3). With our much enlarged sample of quasars at $z>5.6$,  we have an opportunity to revisit this issue by creating composite spectra of our quasar sample as well as from subsamples with different emission line properties.

\begin{figure}[ht]
\centering
\includegraphics[scale=0.59]{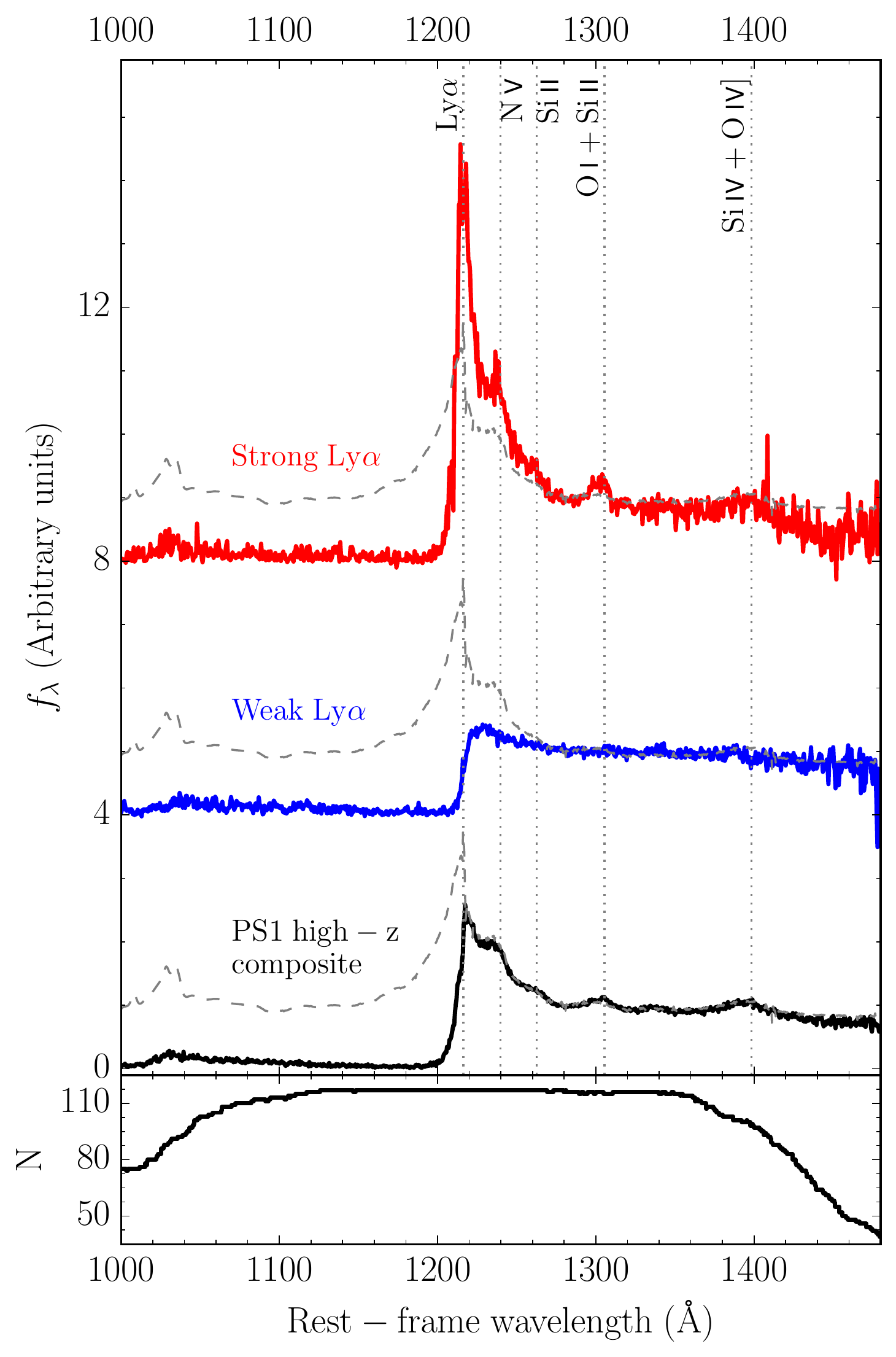}
\caption{
\textit{Top:} The arithmetic median composite spectrum of 117 quasars belonging to the PS1 sample is shown in black. The gray dashed line is the low-redshift composite quasar spectrum from \cite{selsing2016} for comparison. The blue (red) line is the composite spectrum determined from 10\% of individual spectra with the weakest (strongest) $\lyaew$. The blue and red spectra are vertically shifted by 4 and 8 units, respectively. 
\textit{Bottom:} Number of quasars per wavelength bin contributing to the PS1 high-redshift quasar composite spectrum (black line in the top panel).
 These composite spectra show the diversity of the PS1 quasars in terms of their emission line properties. The $z>5.6$ PS1 quasar composite spectrum is available from the online journal in Table \ref{tab:composite}.
}
\label{fig:composite}
 \end{figure}

The black line in Figure \ref{fig:composite} shows the composite spectrum of 117 $z>5.6$ quasars from the PS1 sample. These spectra include all 77 PS1-discovered quasars plus other spectra kindly provided by the authors of their discovery papers. 
In order to create the composite, we first normalized every individual spectrum to its median flux in the rest-frame wavelength range 1285--1295\,\AA, which is a region free of emission lines \citep[e.g.,][]{vandenberk2001}. Next, we resampled the spectra to a common wavelength grid and combined them using a simple arithmetic median, which preserves the relative fluxes of emission lines \citep{vandenberk2001}.  
 As can be seen from Figure \ref{fig:composite},  redward of Ly$\alpha$ the composite spectrum agrees fairly well with the low-redshift composite spectrum of bright ($r\lesssim 17$) $1.0 < z< 2.1$ quasars from \cite{selsing2016} (gray dashed line). Blueward of Ly$\alpha$ the emission is virtually zero due to the strong IGM absorption at these redshifts.

 We created two additional composite spectra following the same procedure as above. One from the  10\% of spectra with the largest  rest-frame  $\lya + \nv$ equivalent width ($\lyaew$) values and the other from the 10\% of spectra with the smallest  $\lyaew$\ values.
In order to estimate the EWs of our quasar sample, we follow the procedure of \cite{diamond2009}. In short, we fit a power law of the form $f_\lambda=C\times \lambda^\beta$ to regions usually free of emission lines (1285--1295, 1315--1325, 1340--1375, 1425--1470, 1680--1710, 1975--2050, and 2150--2250 \AA). We then obtain the EW by integrating the flux above the continuum between $\lambda_{\rm rest} = 1160\,$\AA\ and $\lambda_{\rm rest} = 1290\,$\AA. However, as discussed in Section \ref{sec:m1450}, it is challenging to obtain a robust fit of the continuum for most of our spectra. Therefore, to circumvent this difficulty, we fix the power law index $\beta$. We estimate the EWs using two different power-law indices: $\beta=-1.5$ \citep{vandenberk2001} and $\beta=-1.7$ \citep{selsing2016}. Finally, we average these EWs.

 The two composite spectra with the strong and weak emissions are shown in Figure \ref{fig:composite} with red and blue lines, respectively. 
The differences are quite evident, on the one hand the first composite spectrum shows a Ly$\alpha$ line much stronger than the composite of low-redshift bright quasars, while the Ly$\alpha$ line in our second composite spectrum is virtually absent and resembles a weak-emission line quasar (see Section \ref{sec:wlq}). We note that, in general, redshift uncertainties are larger for weak-lined objects, which might blur out emission features in a composite spectrum even more. 
The mean redshift and $M_{1450}$ of the quasars used for the composite spectra with strong and weak emissions are ($6.05 \pm 0.11$; $-26.28 \pm 0.49$) and ($5.91 \pm 0.18$; $-27.03 \pm 0.88$), respectively.

The three composite spectra created in this section are available in Table \ref{tab:composite}.

\section{Weak emission line quasars}
\label{sec:wlq}
\cite{diamond2009} studied a sample of $\sim3000$ quasars in the redshift range $3<z<5$. They found that the distribution of $\lyaew$ follows a log-normal distribution, and defined weak-line quasars as the 3$\sigma$ outliers at the low-end of this distribution, i.e., $\mathrm{EW}<15.4$\,\AA. This study also showed that the fraction of weak-line quasars evolves with redshift, increasing from 1.3\% at $z<4.2$ to $6.2\%$ at $z>4.2$.

It has been argued that  the fraction of weak-line quasars at $z\sim 6$ could reach $\sim 25\%$, i.e., they are much more abundant than the 1--6\% fraction observed at lower redshifts \citep{banados2014}. 
Several scenarios have been proposed to explain the existence of these intriguing weak-line quasars, but no consensus has been reached (e.g., \citealt{laor2011,wu_j2012,wang_jm2014,luo2015,shemmer2015}). A large fraction of weak-line quasars at the highest accessible redshifts could support evolutionary scenarios suggesting that these rare quasars may be in such an early formation phase that their broad-line region is not yet in place \citep{liu2011}.  We will revisit this issue with the 117 $z>5.6$ PS1 quasars used in Section \ref{sec:composite}.   

\begin{figure}[ht]
\centering
\includegraphics[scale=0.38]{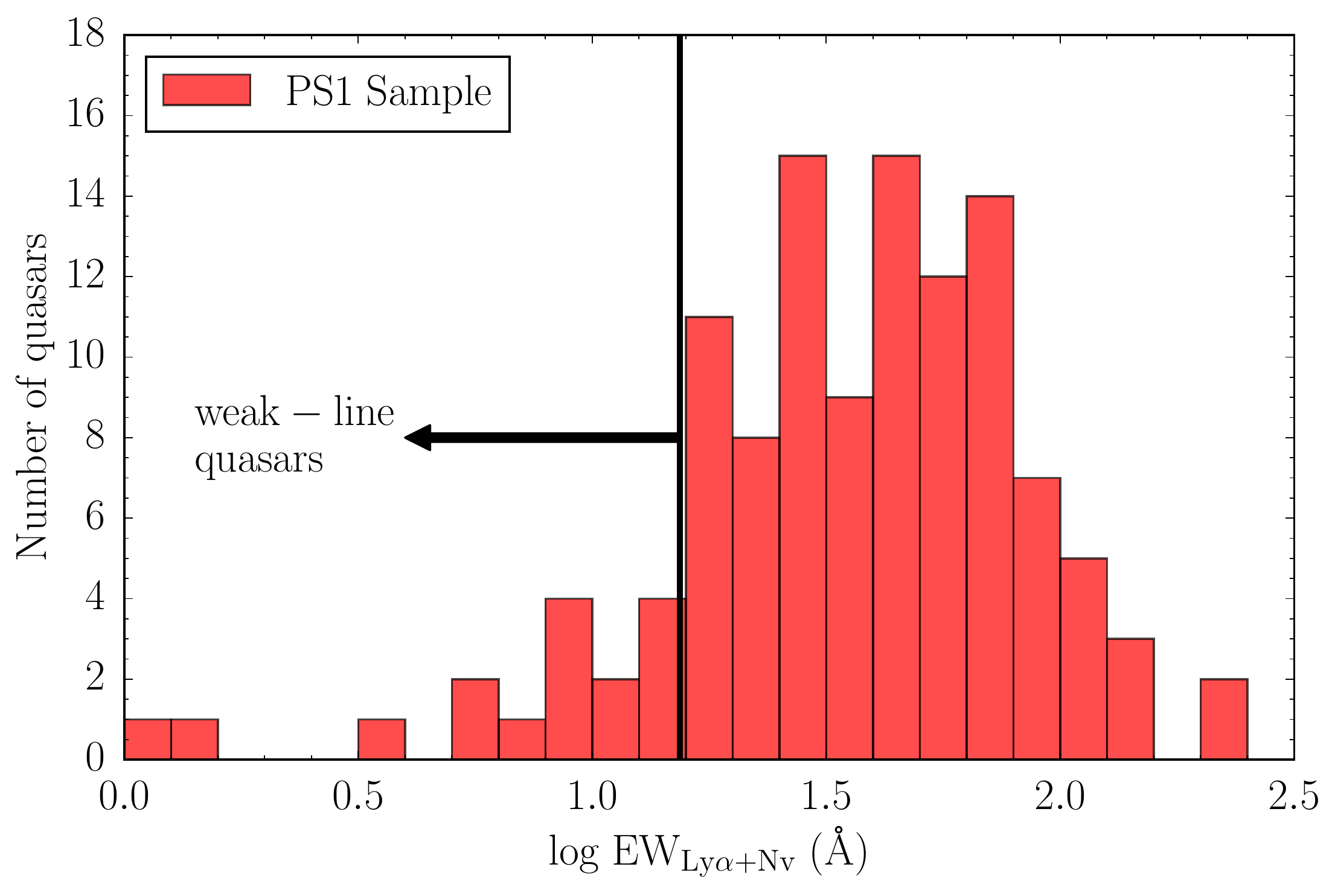}
\caption{
Distribution of rest-frame $\lyaew$\ for 117 $z>5.6$ quasars from the PS1 sample (see Section \ref{sec:wlq}). Using the definition of weak-line quasars at $\lyaew<15.4$\,\AA\, \citep{diamond2009}, 13.7\% (16/117) of the PS1 quasars fall in this category. 
}
\label{fig:wlq}
 \end{figure}

 Figure \ref{fig:wlq} shows the distribution of $\lyaew$\ of the PS1 sample, as estimated in Section \ref{sec:composite}. 13.7\% (16/117) of the quasars satisfy the weak-line quasar definition of \cite{diamond2009}. Even though this fraction is larger than what is found at lower redshifts, it is significantly lower than our initial discoveries suggested \citep{banados2014}. 

\begin{figure}[ht]
\centering
\includegraphics[scale=0.38]{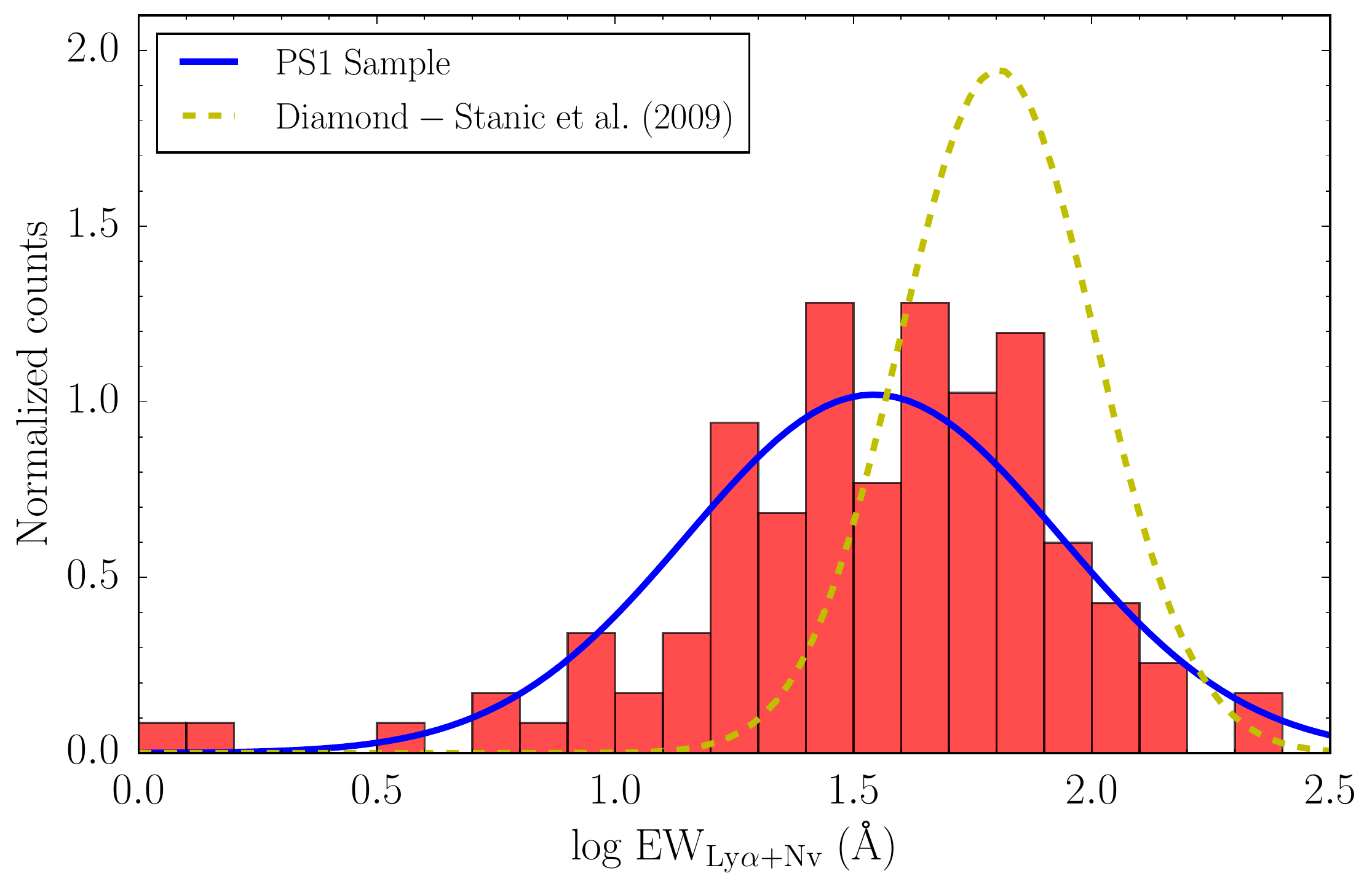}
\caption{
Normalized counts of the $\lyaew$\ distribution from Figure \ref{fig:wlq}. The blue line shows the best-fit log-normal distribution to the data, with $\langle  \log \mathrm{EW (\AA)} \rangle = 1.542$ and $\sigma(\log \mathrm{EW (\AA)}) = 0.391$. The dashed yellow line is the best-fit found at lower redshift by \cite{diamond2009}, with $\langle  \log \mathrm{EW (\AA)} \rangle = 1.803$ and $\sigma(\log \mathrm{EW (\AA)}) = 0.205$. 
 The PS1 sample distribution is systematically shifted to smaller EWs and has a larger dispersion. This could be due to the stronger IGM absorption at $z>5.6$. Alternatively, this might be an indication of a change of the EW distribution with redshift. 
}
\label{fig:wlq-dist}
 \end{figure}

In Figure \ref{fig:wlq-dist} we show the best-fit log-normal distribution to our data, with $\langle  \log \mathrm{EW (\AA)} \rangle = 1.542$ and $\sigma(\log \mathrm{EW (\AA)}) = 0.391$ (blue line). In comparison with the best-fit found by \cite{diamond2009} (yellow dashed line), our best-fit distribution peaks at lower EWs and has a larger dispersion. These two effects could be explained due to the increased opacity of the IGM at $z>5.6$. The $\lya$\ line at $z>5.6$ is, on average, more absorbed than in the quasars studied by \cite{diamond2009}. 
As a simple test, we measure the $\lyaew$ of the original \cite{vandenberk2001} and \cite{selsing2016} templates and of modified versions where all the flux shortward of $\lya$ is set to zero. The differences between the two versions are $15.0 \pm 5.8\,$\AA\ and $26.8 \pm 6.74\,$\AA\ for the \cite{vandenberk2001}  and \cite{selsing2016} templates, respectively. These differences are comparable with the overall mean shift of $28.7\,$\AA\ found between the $\lyaew$ distributions of Figure \ref{fig:wlq-dist}. 
Alternatively, it could be that the EW distributions are actually different, in which case we would need to reconsider the definition of a weak-line quasar. For instance, the lower 3$\sigma$ cut of our best-fit distribution is $<2.3\,$\AA, which is very different to the $15.4\,$\AA\ found by \cite{diamond2009}. It would be interesting to test whether there is an evolution of $\lyaew$ distribution with redshift, in addition to the weak-line fraction evolution. 

\begin{figure}[ht]
\centering
\includegraphics[scale=0.58]{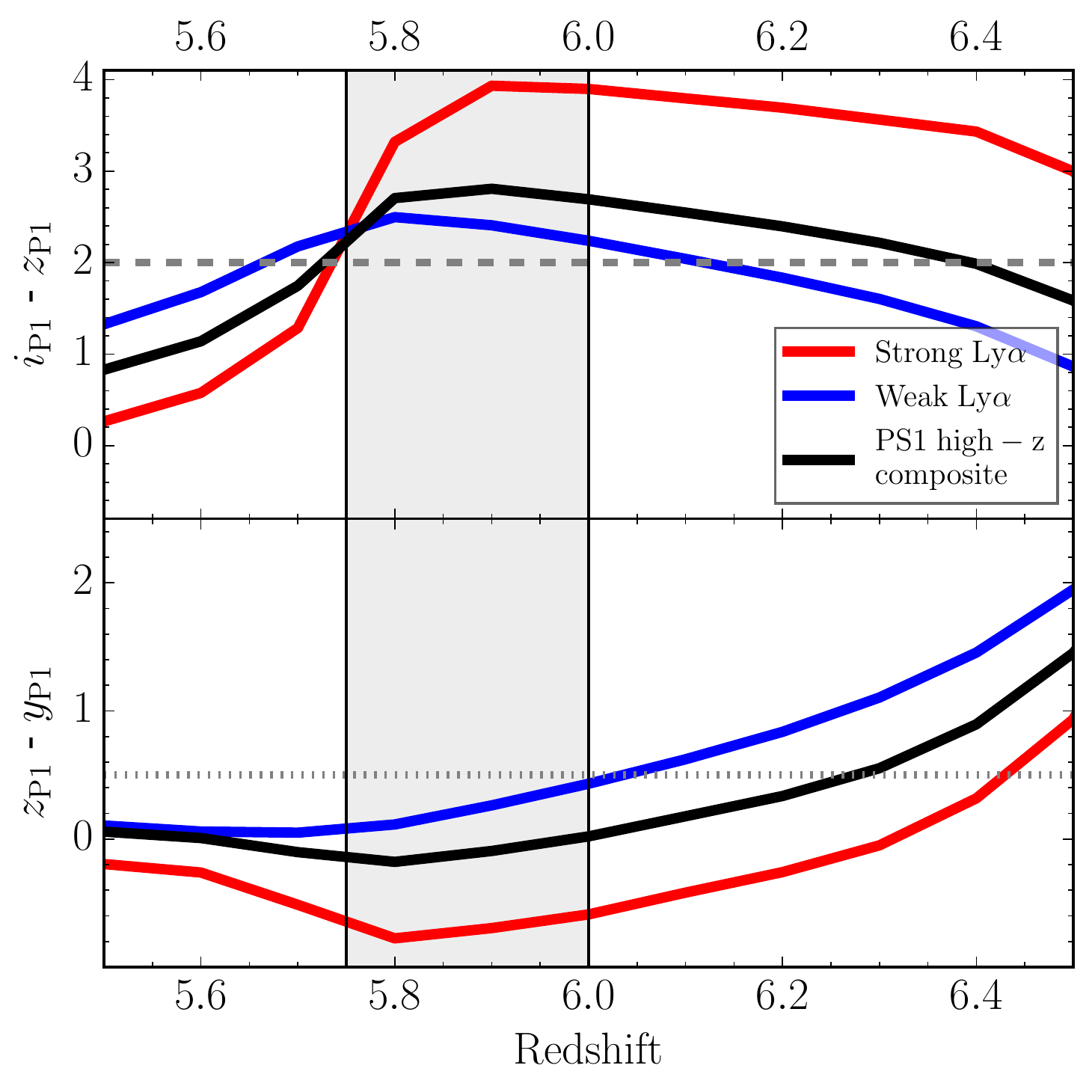}
\caption{
Redshift vs. $\ips-\zps$ (top) and $\zps - \yps$ (bottom) colors. 
The red, blue, and black solid lines are the color tracks of the composite spectra from Section \ref{sec:composite} (see legend and Figure \ref{fig:composite}) redshifted from $z=5.5$ to $z=6.5$. The gray dashed line shows our selection criteria $\ips - \zps > 2$, while the dotted line at $\zps-\yps=0.5$ shows the boundary for which we have different selection criteria (Section \ref{sec:idrops1}). Quasars with weak $\lya$ line at $z>6.2$ are difficult to identify with our current color selection. The shaded area at $5.75<z<6.00$ represents the region where we are sensitive to select quasars with both strong and weak $\lya$ emission lines.
}
\label{fig:wlq-bias}
 \end{figure}

An important point is the impact of selection effects, since quasar color selection is significantly affected by the strength of the $\lya$\ emission line. Figure \ref{fig:wlq-bias} shows the redshift vs. $\ips-\zps$ and $\zps - \yps$ colors tracks for the composite spectra of Section \ref{sec:composite}. It is clear that at $5.65<z<5.75$ we are biased to find more quasars with weaker 
$\lya$, while at $z>6.2$ these weak-line quasars are hard to select. Coincidentally, the redshift range  $5.75<z<6.00$ is where we are sensitive to select both quasars with strong and weak $\lya$\ and also corresponds to the color selection region where our follow-up is more complete. The weak-line fraction of quasars in this redshift range is 12.1\% (7/58), consistent with the estimate based on the whole sample considering Poisson errors. We note that quasars with weak $\lya$ are in general closer to our selection boundaries and thus more prone to be missed in comparison to quasars with normal or strong $\lya$. For example, the $z=5.71$ quasar J1621+5155 with $\lyaew < 5\,$\AA\ \citep{wang2008b} is not part of the PS1 distant quasar sample only due to its $\ips -\zps = 1.85 \pm 0.07$ color. 
Therefore, quasars with weak $\lya$\ line have higher incompleteness in our sample and, as a consequence, it is likely that the fraction of weak-line quasars is underestimated.

Our current high-redshift sample seems to confirm a larger fraction of quasars with weak emission lines, if we use the definition of \cite{diamond2009}. However, we note that the EW distribution we find is significantly different to the one reported by \cite{diamond2009} (Figure \ref{fig:wlq-dist}), which might have consequences on what we call a weak-line quasar. We caution that the results of this section are mainly based on the Ly$\alpha$ line, which is complicated and particularly challenging to interpret at the end of cosmic reionization. It is then critical to test the fraction of weak-line quasars studying other strong broad emission lines such as \civ\ and \mgii\  (see e.g., \citealt{plotkin2015}) in order to make sure that the evolution we see is not only due to the increase of the IGM neutral fraction with redshift. We will therefore postpone a more thorough analysis of the fraction of weak-line quasars until we obtain near-infrared spectroscopy for a representative sample of our quasars. This effort is currently underway.

\section{Summary and Conclusions}
\label{sec:summary}
One of the key challenges of modern astronomy is to study and understand the earliest sources and structures of the universe---their formation and evolution across cosmic time. Luminous quasars at the highest accessible redshifts are ideal tools to probe the early universe. However, strong conclusions from their study have previously been limited by low number statistics.

 In this paper, we have described our method to identify quasars in the redshift range $5.6 \lesssim z \lesssim 6.5$ by mining the Pan-STARRS1 database (Section \ref{sec:ps1-selection}) complemented by follow-up optical and near-infrared observations (Section \ref{sec:ps1-followup}). This is an update to the criteria presented in \cite{banados2014} and complements our method to find quasars at $z\gtrsim 6.5$ described in \cite{venemans2015}. 
In total,  we have so far discovered 77 quasars at  $5.6 \lesssim z \lesssim 6.7$ (63 new discoveries presented in this paper; see Section \ref{sec:ps1-discoveries}), almost doubling the number of quasars previously known within the first gigayear of the universe (see Figure \ref{fig:qso-spectra}). It is important to note that a large fraction of these newly discovered quasars are in the southern sky (see Figure \ref{fig:sky-distribution}), which 
constitutes the ideal ground for follow-up investigations using facilities such as ALMA, VLT, and the Magellan Telescopes. 

Our search is still on-going and now using the latest data release of the Pan-STARRS1 survey, which was recently made available. In the short term, we plan to mine this deeper dataset to complete a more homogeneous quasar sample at $5.75<z<6.00$, which can be used to provide an updated $z\sim 6$ quasar luminosity function, more accurate than what was possible with previous, smaller quasar samples. We are currently working to understand and model the selection function and completeness of our survey. The challenge is to take into account the inhomogeneous depth of different bands across the sky (Figure \ref{fig:limmag}) and the big impact that diverse emission line properties have on our selection (Figure \ref{fig:wlq-bias}), thus rigorous modeling is needed.  We will eventually use this modeling work to constrain the quasar luminosity function also at higher redshifts, where the discovery of more quasars is imminent.

In Section \ref{sec:ps1-sample}, we have introduced the PS1 distant quasar sample which currently consists of 124 quasars that satisfy our selection criteria, while a complete census of all quasars currently known at $z>5.6$ is provided in the Appendix \ref{append:qsolist}. Composite spectra of the PS1 quasar sample are presented in Section \ref{sec:composite}. The PS1 distant quasar sample spans a large range of redshifts and luminosities (see Figure \ref{fig:z-uv}) and shows a broad variety of spectral features, including quasars with 
very strong Ly$\alpha$ emission and others with weak or completely absorbed Ly$\alpha$ (see Figures \ref{fig:qso-spectra} and \ref{fig:composite}).

 We revisited the issue of how common weak emission line quasars are at high-redshift in Section \ref{sec:wlq}. Following \cite{diamond2009}, we find that $13.7\%$ of our quasars are classified as weak-line quasars (Figure \ref{fig:wlq}), a larger fraction than what is found at lower redshifts. However, we note that these results are based mostly on $\lya$\ and the weak-line classification for some of these quasars is not irrefutable (see Fig. \ref{fig:wlq-dist}). 
 Therefore, near-infrared spectroscopy of this sample is required to establish whether some of these objects are real weak-line quasars or their Ly$\alpha$ is being significantly affected by the neutral IGM (see e.g., Figure 3 in \citealt{banados2014}).

So far, most of the studies of high-redshift quasars have focused on individual objects or a few sources. The coming years should see a vast improvement of our understanding of the early universe through studies of the distant quasar population presented in this work. This will require a combined and dedicated effort of X-rays, optical, (near-)infrared, (sub-)millimeter, and radio observations using the current and next generation of ground and space based telescopes. 
Moreover, the upcoming surveys and facilities will enable us to find fainter quasars and push the redshift frontier even further. 
To conclude, 
the large number of quasars presented in this paper is merely the start of an exciting transition era towards  a statistical characterization of the earliest massive black holes and galaxies in the universe.

\acknowledgments

We acknowledge the following people who assisted in some of the
observations presented herein:
Roberto Assef, 
Carla Fuentes, 
David Girou, 
Fiona Harrison,
Bing Jiang, 
George Lansbury, 
Elena Manjavacas,  
Alejandra Melo, 
Ga\"el Noirot, 
Michael Rauch,
Eden Stern. 
We also thank Carlos Contrerars for his support with the data reduction of the near-infrared images taken with Retrocam at the du Pont Telescope. 
E.P.F. and B.P.V. acknowledge funding through the ERC grant `Cosmic Dawn'. 
X.F., I.M., and JT.S. acknowledge support from NSF grant AST 11-07682 and 15-15115.
E.S. acknowledges support for this work provided by NASA through Hubble
Fellowship grant HST-HF2-51367.001-A awarded by the Space Telescope
Science Institute, which is operated by the Association of
Universities for Research in Astronomy, Inc., for NASA, under contract
NAS 5-26555.

The Pan-STARRS1 Surveys (PS1) have been made possible through contributions of the Institute for Astronomy, the University of Hawaii, the Pan-STARRS Project Office, the Max-Planck Society and its participating institutes, the Max Planck Institute for Astronomy, Heidelberg and the Max Planck Institute for Extraterrestrial Physics, Garching, The Johns Hopkins University, Durham University, the University of Edinburgh, Queen's University Belfast, the Harvard-Smithsonian Center for Astrophysics, the Las Cumbres Observatory Global Telescope Network Incorporated, the National Central University of Taiwan, the Space Telescope Science Institute, the National Aeronautics and Space Administration under grant No. NNX08AR22G issued through the Planetary Science Division of the NASA Science Mission Directorate, the National Science Foundation under grant No. AST-1238877, the University of Maryland, and Eotvos Lorand University (ELTE).

This work is based on observations made with ESO Telescopes at the La Silla Paranal Observatory under programs 091.A-0421, 092.A-0150, 092.A-0339, 093.A-0574, 093.A-0863, 094.A-0053, 094.A-0079, 095.A-0375, 095.A-0535, 096.A-0291, 096.A-0420, 381.A-0486,  
and ID 179.A-2010 (PI. McMahon).

This paper includes data gathered with the 6.5 meter Magellan Telescopes located at Las Campanas Observatory, Chile. The FIRE observations were supported by the NSF under grant AST-1109915.

Some of the data presented herein were obtained at the W.M. Keck Observatory, which is operated as a scientific partnership among the California Institute of Technology, the University of California and the National Aeronautics and Space Administration. The Observatory was made possible by the generous financial support of the W.M. Keck Foundation. 
The authors wish to recognize and acknowledge the very significant cultural role and reverence that the summit of Mauna Kea has always had within the indigenous Hawaiian community.  We are most fortunate to have the opportunity to conduct observations from this mountain.

Based on observations collected at the Centro Astron\'omico Hispano Alem\'an (CAHA) at Calar Alto,
operated jointly by the Max-Planck Institut f\"ur Astronomie and the Instituto de Astrof\'isica de Andaluc\'ia (CSIC)

Observations reported here were obtained at the MMT Observatory, a joint facility of the University of Arizona and the Smithsonian Institution.

The LBT is an international collaboration among institutions in the United States, Italy and Germany. The LBT Corporation partners are: The University of Arizona on behalf of the Arizona university system; Istituto Nazionale di Astrofisica, Italy;  LBT Beteiligungsgesellschaft, Germany, representing the Max Planck Society, the Astrophysical Institute Potsdam, and Heidelberg University; The Ohio State University; The Research Corporation, on behalf of The University of Notre Dame, University of Minnesota and University of Virginia.
%
This paper used data obtained with the MODS spectrograph built with
funding from NSF grant AST-9987045 and the NSF Telescope System
Instrumentation Program (TSIP), with additional funds from the Ohio
Board of Regents and the Ohio State University Office of Research.
%

Part of the funding for GROND (both hardware as well as personnel) was 
  generously granted from the Leibniz-Prize to Prof. G. Hasinger 
  (DFG grant HA 1850/28-1).

UKIDSS and UHS use the UKIRT Wide Field Camera (WFCAM; \citealt{casali2007}) and a photometric system described in \cite{hewett2006}. The science archive is described in \cite{hambly2008}.

This publication makes use of data products from the \textit{Wide-field Infrared Survey Explorer},
which is a joint project of the University of California, Los Angeles, and the Jet Propulsion Laboratory/California Institute of Technology,
funded by the National Aeronautics and Space Administration.

This research has benefitted from the SpeX Prism Spectral Libraries, maintained by Adam Burgasser at \url{http://pono.ucsd.edu/~adam/browndwarfs/spexprism}

This research made use of Astropy, a community-developed core Python package for Astronomy \citep[][\url{http://www.astropy.org}]{astropy13}.
The plots in this publication 
were produced using Matplotlib \citep[][\url{http://www.matplotlib.org}]{hunter2007}.

{\it Facilities:} \facility{PS1 (GPC1)}, \facility{VLT:Antu (FORS2)}, \facility{NTT (EFOSC2)}, \facility{LBT (MODS)}, \facility{Max Planck:2.2m (GROND)},
 \facility{Magellan:Baade (FIRE)}, \facility{Magellan:Clay (LDSS3)}, \facility{Keck:I (LRIS)}, 
\facility{Hale (DBSP)}, 
\facility{CAO:3.5m (Omega2000)}, \facility{CAO:2.2m (CAFOS)},
 \facility{MMT (SWIRC)}, \facility{Du Pont (Retrocam)}

\appendix

\section{List of quasars at $\lowercase{z}>5.6$} \label{append:qsolist}

Table \ref{tab:qsos-info} lists the full names, coordinates, redshifts, and rest-frame 1450\,\AA\ magnitudes 
for the 173 $z>5.6$ quasars known as of the end of 2016 March. The table column ``PS1'' is $>0$ for quasars that satisfy the selection criteria presented in Section \ref{sec:ps1-selection} and \cite{venemans2015}. There are 124 PS1-selected quasars, out of which 77 are PS1 discoveries.

\section{Pan-STARRS1, $J$, and \textit{WISE} photometry of the known $\lowercase{z}>5.6$ quasars} \label{append:photometry}

\begin{figure}[ht]
\centering
\includegraphics[scale=0.75]{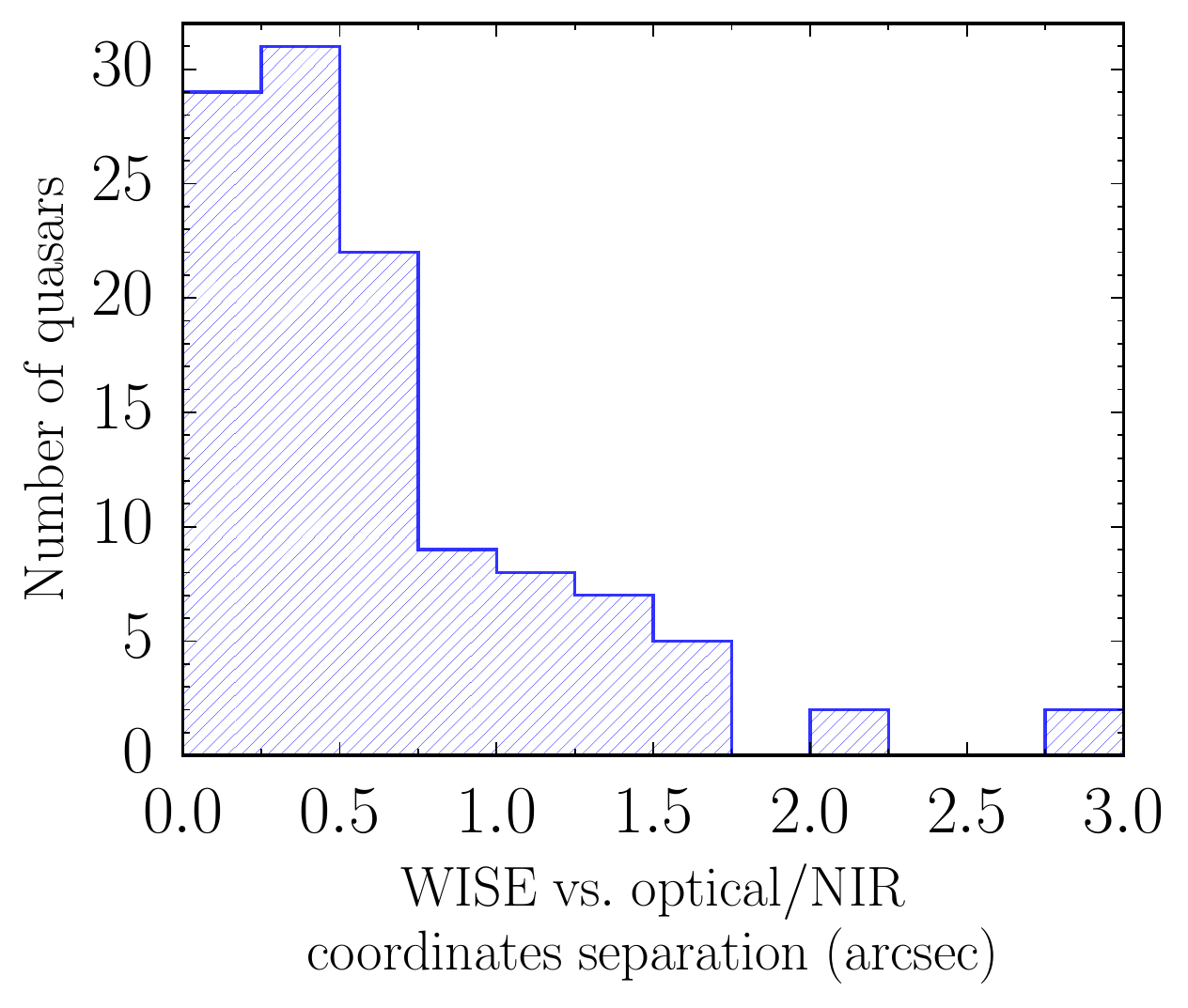}
\caption{
Separation in arcsec between the \textit{WISE} and optical/near-infrared coordinates of the known $z>5.6$ quasars with at least a 3$\sigma$ detection in one of the \textit{WISE} bands.  The four quasars with a separation greater than 2\arcsec are: P135+16, P187--02,  P242--12, and P197+25.
}
\label{fig:wisesep}
 \end{figure}

The quasars were cross-matched to the 
\textit{WISE} All-Sky data release products catalog \citep{cutri2012} and to the ALLWISE Source catalog 
and Reject Table\footnote{\url{http://wise2.ipac.caltech.edu/docs/release/allwise/expsup/sec2_1.html}} \citep{cutri2014} within a radius of 3\arcsec\ (note that the \textit{WISE} PSF FWHM for $W1$ and $W2$ is about 6\arcsec). 
Figure \ref{fig:wisesep} shows a histogram with the separation between the optical/near-infrared and \textit{WISE} positions. Most of \textit{WISE} coordinates are within 2\arcsec\ from the optical/near-infrared locations. The four objects with a separation greater than 2\arcsec\ are P135+16, P187--02, P242--12, and P197+25, with separations of $2.95\arcsec$, $2.83\arcsec$, $2.11\arcsec$, and $2.02\arcsec$, respectively. Therefore, their \textit{WISE} magnitudes must be used with caution. As a test we can compare the $W1$ magnitudes of three of these objects with their  \textit{Spitzer} 3.6$\,\mu m$ photometry, taken with our \textit{Spitzer} survey of our PS1 quasar sample (Program: 11030; PI: R.~Decarli; details of the survey will be presented by R.~Decarli in prep.). The $W1$ magnitudes for P135+16, P187--02, and P242--12 are $19.51\pm 0.11$, $19.89\pm0.14$, and $19.00 \pm 0.07$,  while their  \textit{Spitzer} 3.6$\mu m$ magnitudes are $20.49\pm 0.02$, $20.38 \pm 0.01$, and $19.30 \pm 0.01$. The magnitudes are significantly different, especially for P135+16 and P187--02, the objects with the largest separations. P135+16 is a radio-loud quasar with a radio-loudness parameter of $R=91.4\pm 8.8$ \citep{banados2015a}. However, if the  \textit{Spitzer} photometry is used as a proxy of the optical luminosity instead of the \textit{WISE} photometry, its radio-loudness parameter increases to $R=229.9 \pm  11.7$, making it the radio-loudest quasar at $z>5.5$ (see Figure 2 in \citealt{banados2015a}). 
Based on Figure \ref{fig:wisesep} and the discrepancies between the \textit{WISE} and  \textit{Spitzer} photometry we only include in Figure~\ref{fig:wise-crit} objects with a separation between the \textit{WISE} and optical/near-infrared coordinates smaller than 2\arcsec.

\begin{figure}[ht]
\centering
\includegraphics[scale=0.75]{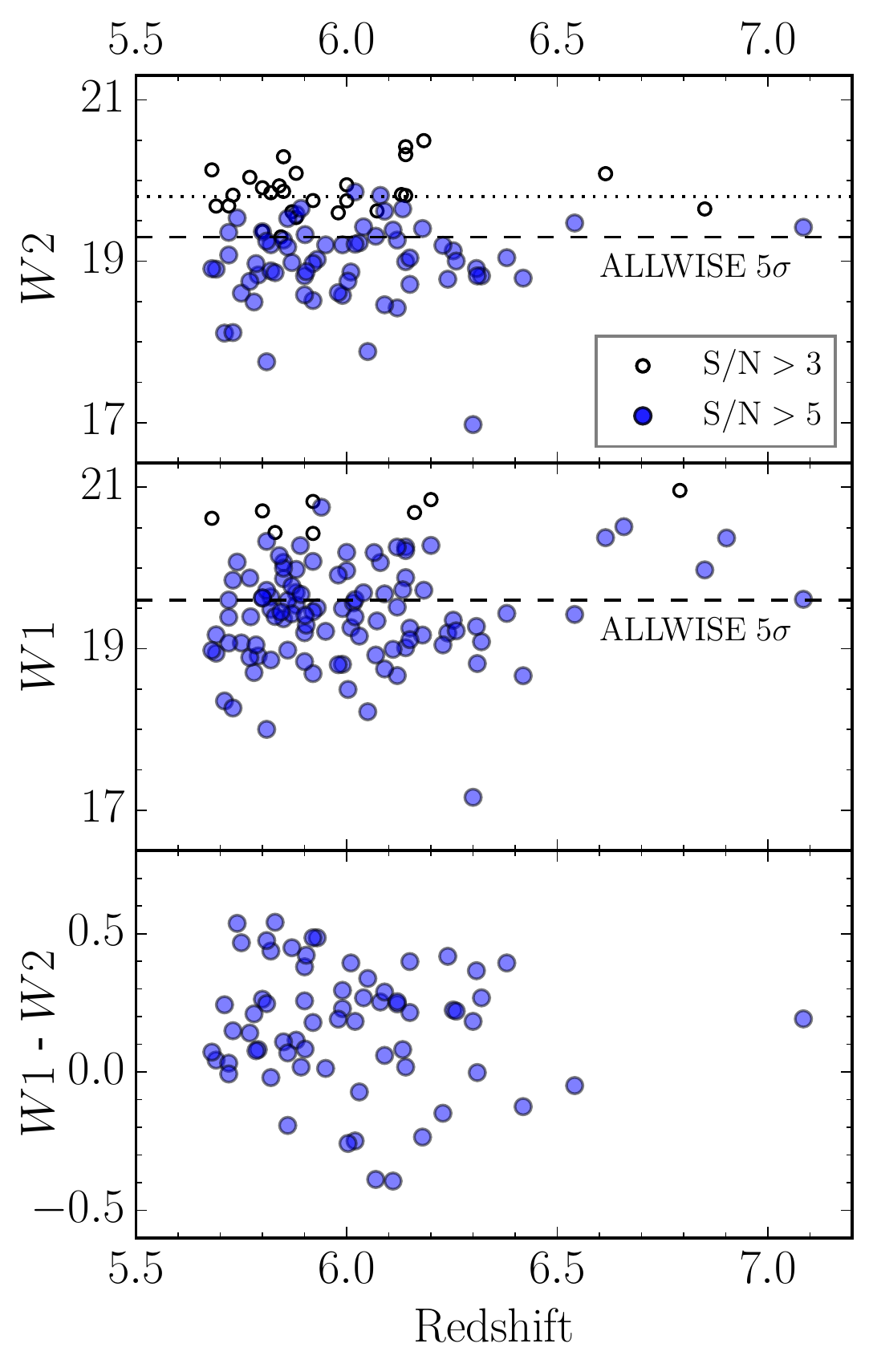}
\caption{
 \textit{WISE} $W1$ (middle) and $W2$ (top) magnitudes vs. redshift for all $z>5.6$ quasars detected in at least one of the \textit{WISE} bands with a S/N$>3$. The bottom panel shows the $W1 - W2$ color vs. redshift for quasars with S/N$>5$ in both bands. The dashed lines represent the nominal ALLWISE $5\sigma$ limiting magnitudes, while the dotted line in the top panel represents the expected median limiting magnitude once ALLWISE is combined with the ongoing NEOWISE reactivated program. 
}
\label{fig:wise-redshift}
 \end{figure}

Figure \ref{fig:wise-redshift} shows the \textit{WISE} magnitudes and colors for the 111 ($\sim 65\%$) known $z>5.6$ quasars that are detected in at least one of the \textit{WISE} bands with a S/N\,$>3$. The dashed lines show the nominal $5\,\sigma$ ALLWISE limiting magnitudes but there are region of the sky that reach significantly deeper magnitudes. The dotted line represents the expected median $5\,\sigma$ limiting magnitude once the data from ALLWISE is combined with the on-going  NEOWISE reactivated mission \citep{mainzer2014}, which is surveying the sky in $W1$ and $W2$ to search for near-Earth objects. This increase in depth will also be beneficial to efficiently search for the brightest quasars at the highest accessible redshifts ($\gtrsim 7$).

Table \ref{tab:qsos-ps1wiseinfo} presents the \textit{WISE}, $J$-band, PS1 PV3 magnitudes of the 173 $z>5.6$ quasars known by March 2016. 
Almost 81\% (140/173) of the known  $z>5.6$ quasars are detected in at least one of the PS1 bands with a S/N$>5$, while a similar fraction (141/173) have $J$-band information with S/N\,$>5$. There 
are seven known quasars with Decl.$<-30\degr$, which are 
therefore outside the PS1 footprint.

\section{Spectroscopically rejected candidates} \label{append:rejected}

Table \ref{tab:spc-rejected} presents the 11 candidates whose spectra showed they were not $z>5.6$ quasars. Objects with PS1 column equals 1 are mostly M and L dwarfs, while sources with PS1 column 2 or 3 are more likely late L-dwarfs and T-dwarfs. However, a thorough classification of these objects is beyond the scope of this work.


\clearpage

\LongTables
\begin{deluxetable*}{lcl}
\tablecolumns{3}
    \tablewidth{0pt}
    
\tablecaption{Follow-up photometry of the PS1-discovered quasars selected as $i$-dropouts with $\ips - \zps < 0.5$ ($5.7 \lesssim z \lesssim 6.2$) \label{tab:follow-idrops1}}
\tablehead{
\colhead{Quasar} & \colhead{$z$} & \colhead{Photometry (mag)}
}
\startdata
P002--06 & 5.93 & $\igrond>23.01$ \; $\zgrond=20.55 \pm 0.05$ \; $\Jgrond=19.86 \pm 0.14$ \; $\Hgrond=19.92 \pm 0.20$\\
         &      & $\intt= 21.72 \pm 0.09$  \\
P002+32 & 6.10  & $\Icafos>23.30$ \; $\zotk=21.65 \pm 0.09$ \;  $\Jotk=21.38 \pm 0.19$ \\
P004+17 & 5.80 &  $\Icafos=23.34 \pm 0.35$ \; $\zotk=21.14 \pm 0.14$ \; $\Yotk=20.79 \pm 0.22$ \; $\Jotk=20.67 \pm 0.16$  \\
P004--24 & 5.68 & $\igrond=22.09 \pm 0.15$ \; $\zgrond=19.66 \pm 0.03$ \; $\Jgrond=19.30 \pm 0.08$ \; $\Hgrond=18.97 \pm 0.08$ \\
         &      & $\intt=20.81 \pm 0.03$ \; $\zntt=19.56 \pm 0.04$  \\
P007+04 & 6.00 & $\igrond=22.74 \pm 0.24$ \; $\zgrond=20.39 \pm 0.04$ \; $\Jgrond=19.77 \pm 0.11$ \; $\Hgrond=19.71 \pm 0.15$  \\
P009--08 & 5.72 & $\igrond=22.24 \pm 0.10$ \; $\zgrond=20.23 \pm 0.02$ \; $\Jgrond=19.79 \pm 0.08$ \; $\Hgrond=19.70 \pm 0.13$\\
\enddata
\tablecomments{Quasars sorted by right ascension. Their complete names and coordinates are given in Table \ref{tab:qsos-info}. Their PS1 magnitudes are listed in Table \ref{tab:qsos-ps1wiseinfo}. A machine readable format of this table is available in the online journal.
       }
\end{deluxetable*}

\clearpage
\begin{deluxetable*}{lcl}
    \tablewidth{0pc}
    
\tablecaption{Follow-up photometry of the PS1-discovered quasars selected as $i$-dropouts with $\ips - \zps \geq 0.5$ ($6.2 \lesssim z \lesssim 6.5$) \label{tab:follow-idrops2}}
\tablehead{
\colhead{Quasar} & \colhead{$z$} & \colhead{Photometry (mag)}
}
\startdata
P009--10 & 5.95 & $\igrond >23.39$ \; $\zgrond=21.11 \pm 0.06$ \; $\Jgrond=19.93 \pm 0.07$ \; $\Hgrond=19.72 \pm 0.11$ \\
P065--26 & 6.14 & $\igrond = 22.70 \pm 0.17$ \; $\zgrond =20.30 \pm 0.03$ \; $\Jgrond =19.32 \pm 0.06$ \; $\Hgrond=19.39 \pm 0.18$\\
         &      & $\intt=22.24 \pm 0.07$  \\
P159--02 & 6.38 & $\intt >23.74$ \; $\Jntt=20.00 \pm 0.10$ \\
P210+27  & 6.14 & $Y=20.93\pm 0.20$\; $J=20.35\pm0.15$\, (UKIDSS)\\
P217--16 & 6.11 & $\intt= 22.53 \pm 0.06$ \; $\zntt=20.44 \pm 0.03$\\
         &      &  $\Yotk>18.43$ \; $\Jotk=19.69 \pm 0.07$ \\
P217--07 & 6.14 & $\igrond=23.58\pm 0.28$ \; $\zgrond=21.05\pm0.03$ \; $\Jgrond=19.62\pm0.09$\\ 
P308--21 & 6.24 & $\igrond>23.60$ \; $\zgrond=20.98 \pm 0.05$ \; $\Jgrond=20.17 \pm 0.11$ \\
P333+26  & 6.03 & $\intt=22.48 \pm 0.17$ \; $\zntt=20.58 \pm 0.07$ \; $\Jntt=20.40 \pm 0.13$ \; $J_{\rm SWIRC}=20.44 \pm 0.05$
\enddata
\tablecomments{Quasars sorted by right ascension. Their complete names and coordinates are given in Table \ref{tab:qsos-info}. Their PS1 magnitudes are listed in Table \ref{tab:qsos-ps1wiseinfo}. 
     Note that even when these quasars were selected with $\zps -\yps \geq 0.5$, most of them have a lower redshift than expected (i.e., $z<6.2$). A machine readable format of this table is available in the online journal.
       }
\end{deluxetable*}

\begin{deluxetable*}{lcl}
    \tablewidth{0pc}
\tablecaption{Follow-up photometry of the PS1-discovered quasars that do not satisfy all our selection criteria \label{tab:follow-extended}}
\tablehead{
\colhead{Quasar} & \colhead{$z$} & \colhead{Photometry (mag)}
}
\startdata
P000+26 & 5.75 & $\zotk=19.49 \pm 0.01$ \; $\Jotk=19.53 \pm 0.02$\\
P089--15 & 6.05 & $\igrond=23.27 \pm 0.20$ \; $\zgrond=19.94 \pm 0.02$ \; $\Jgrond=19.17 \pm 0.04$ \; $\Hgrond=18.47 \pm 0.04$\\
         &      & $\ydp=19.68 \pm 0.04$ \\
P108+08 & 5.92 & $\igrond=22.34 \pm 0.14$ \; $\zgrond=19.35 \pm 0.02$ \; $\Jgrond=19.29 \pm 0.06$ \; $\Hgrond=19.03 \pm 0.07$ \\
P135+16 & 5.63 & $\igrond=22.70 \pm 0.18$ \; $\zgrond=20.85 \pm 0.04$ \; $\Jgrond=20.30 \pm 0.12$ ; $\Hgrond=20.91 \pm 0.33$  \\
P245--00 & 5.68 & $\igrond>23.49$ \; $\zgrond=21.69 \pm 0.09$ \; $\Jgrond=20.78 \pm 0.22$ \\
         &      & $\ydp=21.30 \pm 0.13$ \; $\jdp=21.10 \pm 0.19$ 
\enddata
\tablecomments{Quasars sorted by right ascension. Their complete names and coordinates are given in Table \ref{tab:qsos-info}. Their PS1 magnitudes are listed in Table \ref{tab:qsos-ps1wiseinfo}. 
    These quasars were selected by extended PS1 criteria and they are discussed in 
    Section \ref{sec:individual_notes}. A machine readable format of this table is available in the online journal.
       }
\end{deluxetable*}

\clearpage
\LongTables
\begin{deluxetable*}{lcrrrr}
\tablecolumns{5}
\tablewidth{0pc}
\tablecaption{Spectroscopic observations of the new PS1 quasars \label{tab:spectroscopy}}
 \tablehead{
\colhead{Quasar}    &  \colhead{Date} &   \colhead{Telescope/Instrument}   & \colhead{Exposure Time} &
\colhead{Slit Width} 
}
\startdata
PSO~J000.3401+26.8358   & 2014 Oct 19	      & MMT/Red Channel & 300\,s        & 1\farcs0        \\%
					   & 2015 Nov 5	      &  VLT/FORS2 & 1500\,s        & 1\farcs0        \\%
PSO~J002.1073--06.4345  & 2013 Oct 24        & VLT/FORS2       & 1482\,s        & 1\farcs3 \\ 
PSO~J002.3786+32.8702   & 2014 Dec 23        & P200/DBSP       & 1800\,s       & 1\farcs5 \\ 
PSO~J004.3936+17.0862  & 2016 Feb 3  	      & LBT/MODS        & 2700\,s	 & 1\farcs2	\\
PSO~J004.8140--24.2991   & 2013 Aug 10        & Magellan/FIRE   & 2409\,s       & 0\farcs6       \\ %
			 & 2014 Oct 12	      & VLT/FORS2       & 1452\,s	         & 1\farcs3       \\ %
PSO~J007.0273+04.9571  & 2013 Jul 9	      & VLT/FORS2       & 1782\,s       & 1\farcs3       \\
\enddata
\tablecomments{Quasars sorted by right ascension. For full coordinates and redshifts see Table \ref{tab:qsos-info}. The 14 PS1 quasars published in \cite{morganson2012,banados2014,banados2015a,venemans2015} are included here for completeness.}
\end{deluxetable*}

\begin{deluxetable}{ccccc}
\tablecaption{PS1 distant quasar sample arithmetic median composite spectra (see Figure \ref{fig:composite}). \label{tab:composite}}
\tablehead{\colhead{$\lambda$} & \colhead{$f_{\lambda, all}$} & \colhead{$N$} & \colhead{$f_{\lambda, {\rm strong}}$ } & \colhead{$f_{\lambda, {\rm weak}}$ }\\ \colhead{$\mathrm{\mathring{A}}$} & \colhead{ } & \colhead{ } & \colhead{ } & \colhead{ }}
\startdata
1000.0 & 0.031 & 75 & 0.020 & 0.038 \\
1000.5 & 0.043 & 75 & 0.039 & 0.075 \\
1001.0 & 0.057 & 75 & 0.036 & 0.139 \\
1001.5 & 0.053 & 75 & 0.011 & 0.145 \\
1002.0 & 0.055 & 75 & 0.017 & 0.221 \\
1002.5 & 0.080 & 75 & 0.007 & 0.145 \\
1003.0 & 0.045 & 75 & -0.027 & 0.083 \\
1003.5 & 0.039 & 74 & 0.094 & 0.065 \\
1004.0 & 0.054 & 74 & 0.144 & 0.075 \\
1004.5 & 0.031 & 75 & 0.028 & 0.071
\enddata
\tablecomments{Table \ref{tab:composite} is published
 in its entirety in a machine readable format. A portion is shown here 
 for guidance regarding its form and content.}
\end{deluxetable}

\clearpage
\LongTables
\begin{deluxetable*}{lcccccccr}

    \tablewidth{0pc}
    
\tablecaption{All $z>5.6$ quasars known as of 2016 March. \label{tab:qsos-info}}
\tablehead{\colhead{Quasar} & \colhead{R.A.} & \colhead{Decl.} & \colhead{$z$} & \colhead{Method\,\tablenotemark{a}} & \colhead{$m_{1450}$\,\tablenotemark{b}} & \colhead{$M_{1450}$} & \colhead{PS1\,\tablenotemark{c} } & \colhead{References}\\ \colhead{ } & \colhead{(J2000)} & \colhead{(J2000)} & \colhead{ } & \colhead{ } & \colhead{$\mathrm{mag}$} & \colhead{$\mathrm{mag}$} & \colhead{ } & \colhead{disc./ $z$/ $m_{1450}$}}
\startdata
PSO~J000.3401+26.8358 & 00:01:21.63 & $+$26:50:09.17 & 5.75 & 4 & 19.52 & $-27.16$ & 4 & 1/1/1 \\
SDSS~J0002+2550 & 00:02:39.39 & $+$25:50:34.96 & 5.82 & 3 & 19.39 & $-27.31$ & 1 & 5/22/1 \\
SDSS~J0005--0006 & 00:05:52.34 & $-$00:06:55.80 & 5.85 & 2 & 20.98 & $-25.73$ & 1 & 5/12/1 \\
PSO~J002.1073--06.4345 & 00:08:25.77 & $-$06:26:04.42 & 5.93 & 4 & 20.41 & $-26.32$ & 1 & 1;43/1/1 \\
PSO~J002.3786+32.8702 & 00:09:30.89 & $+$32:52:12.94 & 6.10 & 4 & 21.13 & $-25.65$ & 1 & 1/1/1 \\
PSO~J004.3936+17.0862 & 00:17:34.47 & $+$17:05:10.70 & 5.80 & 4 & 20.69 & $-26.01$ & 1 & 1/1/1 \\
PSO~J004.8140--24.2991 & 00:19:15.38 & $-$24:17:56.98 & 5.68 & 4 & 19.43 & $-27.24$ & 1 & 1/1/1 \\
PSO~J007.0273+04.9571 & 00:28:06.56 & $+$04:57:25.64 & 6.00 & 4 & 20.11 & $-26.64$ & 1 & 36;43/1/1 \\
\enddata
\end{deluxetable*}

\clearpage
\LongTables
\begin{deluxetable*}{lccccccccc}

    \tablewidth{0pc}
    
\tablecaption{PS1, $J$ band, and WISE photometry of the $z>5.6$ quasars known as of 2016 March. \label{tab:qsos-ps1wiseinfo}}
\tablehead{\colhead{Quasar} & \colhead{\ips} & \colhead{\zps} & \colhead{\yps} & \colhead{$W1$} & \colhead{$W2$} & \colhead{$W3$} & \colhead{$J$} & \colhead{$J_{\mathrm{ref}}$} & \colhead{$E(B-V)$}\\ \colhead{ } & \colhead{$\mathrm{mag}$} & \colhead{$\mathrm{mag}$} & \colhead{$\mathrm{mag}$} & \colhead{$\mathrm{mag}$} & \colhead{$\mathrm{mag}$} & \colhead{$\mathrm{mag}$} & \colhead{$\mathrm{mag}$} & \colhead{ } & \colhead{ }}
\startdata
P000+26 & $21.25 \pm 0.04$ & $19.28 \pm 0.02$ & $19.52 \pm 0.04$ & $19.07 \pm 0.07$ & $18.61 \pm 0.11$ & -- & $19.53 \pm 0.02$ & 1 & 0.04 \\
J0002+2550 & $21.82 \pm 0.06$ & $19.07 \pm 0.01$ & $19.40 \pm 0.04$ & $18.86 \pm 0.06$ & $18.88 \pm 0.13$ & -- & $19.42 \pm 0.09$ & 2 & 0.04 \\
J0005--0006 & $23.30 \pm 0.24$ & $20.44 \pm 0.04$ & $20.99 \pm 0.16$ & $20.00 \pm 0.16$ & -- & -- & $20.81 \pm 0.10$ & 7 & 0.03 \\
P002--06 & $22.68 \pm 0.13$ & $20.25 \pm 0.03$ & $20.42 \pm 0.10$ & $19.51 \pm 0.11$ & $19.02 \pm 0.14$ & -- & $19.86 \pm 0.14$ & 1 & 0.03 \\
P002+32 & $>23.82$ & $21.18 \pm 0.06$ & $21.15 \pm 0.15$ & -- & -- & -- & $21.38 \pm 0.19$ & 1 & 0.04 \\
P004+17 & $>23.31$ & $20.82 \pm 0.07$ & $20.70 \pm 0.13$ & $20.71 \pm 0.28$ & $19.91 \pm 0.32$ & -- & $20.67 \pm 0.16$ & 1 & 0.07 \\
P004--24 & $22.31 \pm 0.14$ & $19.62 \pm 0.02$ & $19.43 \pm 0.04$ & $18.98 \pm 0.07$ & $18.91 \pm 0.12$ & $17.30 \pm 0.34$ & $19.30 \pm 0.08$ & 1 & 0.01 \\
P007+04 & $22.84 \pm 0.16$ & $20.56 \pm 0.05$ & $20.12 \pm 0.07$ & $19.97 \pm 0.16$ & $19.75 \pm 0.29$ & -- & $19.77 \pm 0.11$ & 20 & 0.01 \\
\enddata
\tablerefs{(1) This work, (2) UHS, (3) UKIDSS, (4) VHS, (5) \cite{fan2001}, (6) \cite{fan2003}, (7) \cite{fan2004}, (8) \cite{mcgreer2006}, (9) \cite{willott2007}, (10) \cite{jiang2008}, (11) \cite{jiang2009}, (12) \cite{kurk2009}, (13) \cite{mortlock2009}, (14) \cite{willott2009}, (15) \cite{willott2010a}, (16) \cite{willott2010b}, (17) \cite{zeimann2011}, (18) \cite{mcgreer2013}, (19) \cite{venemans2013}, (20) \cite{banados2014}, (21) \cite{banados2015a}, (22) \cite{barnett2015}, (23) \cite{jiang2015}, (24) \cite{kashikawa2015}, (25) \cite{kim_yongjung2015}, (26) \cite{reed2015}, (27) \cite{venemans2015}, (28) \cite{venemans2015b}, (29) \cite{matsuoka2016}}
\tablecomments{Quasars sorted by right ascension, their full names,  coordinates, and redshifts are given in Table \ref{tab:qsos-info}. 
The PS1 magnitudes are \textit{derredened}. Reddened magnitudes can be obtained by adding $\lambda_f \times E(B-V)$ with $\lambda_f = (1.682, 1.322, 1.087)$  for (\ips, \zps, \yps), see \cite{schlafly2011}. 
\textit{WISE} magnitudes are reported when their measurements have S/N$>3$. They are from ALLWISE source catalog when available. 
The magnitudes of J0033--0125, J0109--3040, J0221--0802, J0227--0605, P055--00, P167--13,  P210--12, J1641+3755, J2053+0047, and P328--09 are taken from the ALLWISE reject table. 
The magnitudes of P209--26, P215--16, and P217--16 are taken from the \textit{WISE} All-Sky data release catalog. 
The UKIRT Hemisphere Survey (UHS)\footnote{\url{http://casu.ast.cam.ac.uk/wfcamsp/uhs}} was not used for our selection but here we provide the $J$ band photometry for known quasars detected in the catalog.  
A machine readable format of this table is available in the online journal.}
\end{deluxetable*}

\clearpage
\LongTables
\begin{deluxetable*}{lccccccccc}
    \tablewidth{0pc}
    
\tablecaption{Candidates spectroscopically confirmed to \textit{not} be $z>5.6$ quasars. \label{tab:spc-rejected}}
\tablehead{\colhead{Rejected} & \colhead{R.A.} & \colhead{Decl.} & \colhead{\ips} & \colhead{\zps} & \colhead{\yps} & \colhead{$Y$} & \colhead{$J$} &\colhead{$E(B-V)$} &\colhead{PS1\,\tablenotemark{a}}\\ \colhead{Candidate } & \colhead{J$2000$} & \colhead{J$2000$} & \colhead{$\mathrm{mag}$} & \colhead{$\mathrm{mag}$} & \colhead{$\mathrm{mag}$} & \colhead{$\mathrm{mag}$} & \colhead{$\mathrm{mag}$} & \colhead{$\mathrm{mag}$} & \colhead{ }}
\startdata
P014+37 & 00:56:09.20 & $+$37:34:16.1 & $23.07 \pm 0.24$ & $20.50 \pm 0.05$ & $19.88 \pm 0.06$ & $19.46 \pm 0.18$ & $19.00 \pm 0.08$ & $0.05$ & 2 \\
P023+30 & 01:32:42.94 & $+$30:49:30.1 & $21.87 \pm 0.08$ & $19.14 \pm 0.01$ & $19.39 \pm 0.04$ &  --              &  --              & $0.05$ & 1 \\
P102+55 & 06:49:07.97 & $+$55:35:25.8 & $>23.57$         & $21.07 \pm 0.09$ & $19.25 \pm 0.03$ &  $20.23 \pm 0.06$&  $20.14\pm 0.04$ & $0.06$ & 3 \\
P116--07& 07:44:13.22 & $-$07:46:22.2 & $22.10 \pm 0.19$ & $20.07 \pm 0.04$ & $19.70 \pm 0.06$ &  --              &  $19.43\pm 0.07$ & $0.16$ & 1 \\
P131+25 & 08:47:45.26 & $+$25:47:17.3 & $>23.69$         & $21.39 \pm 0.09$ & $20.67 \pm 0.13$ &  $20.55 \pm 0.16$&  $20.06\pm 0.09$ & $0.04$ & 2 \\
P150+16 & 10:03:39.37 & $+$16:05:31.0 & $23.69 \pm 0.36$ & $21.30 \pm 0.07$ & $20.85 \pm 0.11$ & --               &  $19.90 \pm 0.11$& $0.04$ & 1 \\
P165+25 & 11:02:23.27 & $+$25:39:35.8 & $22.90 \pm 0.23$ & $20.89 \pm 0.06$ & $19.85 \pm 0.07$ & --               &  $19.55 \pm 0.09$& $0.02$ & 2 \\
P215+33 & 14:22:16.12 & $+$33:29:21.3 & $23.19 \pm 0.20$ & $20.55 \pm 0.04$ & $19.56 \pm 0.03$ & --               &  $19.16 \pm 0.09$& $0.01$ & 2 \\
\enddata
\tablenotetext{a}{PS1 criteria: 0=None; 1=Criteria of Section \ref{sec:idrops1} with $\zps -\yps < 0.5 $; 2=Criteria of Section \ref{sec:idrops1} with $\zps -\yps \geq 0.5 $; 3=Criteria of \cite{venemans2015}}
\end{deluxetable*}

\end{document}